\documentclass[apj,numberedappendix]{emulateapj}
\usepackage{apjfonts}  
\usepackage{epsfig}
\usepackage{natbib}
\shorttitle{Supernova feedback efficiency and mass loading in M82}
\shortauthors{Strickland \& Heckman}

\newcommand\pasa{\ref@jnl{PASA}}%
\submitted{Accepted for publication in the Astrophysical Journal}

\newcommand{\eg}{{\rm e.g.\ }}

\newcommand{\ie}{{\rm i.e.\ }}

\newcommand{\etal}{{\rm et al.\thinspace}}

\newcommand{\cm}{{\rm\thinspace cm}}
\newcommand{\km}{{\rm\thinspace km}}
\newcommand{\g}{{\rm\thinspace g}}
\newcommand{\pcc}{\hbox{$\cm^{-3}\,$}}
\newcommand{\gpcc}{\hbox{$\g\cm^{-3}\,$}}
\newcommand{\s}{{\rm\thinspace s}}
\newcommand{\yr}{{\rm\thinspace yr}}
\newcommand{\erg}{{\rm\thinspace erg}}
\newcommand{\ps}{\hbox{\s$^{-1}\,$}}
\newcommand{\pyr}{\hbox{\yr$^{-1}$}}
\newcommand{\ergps}{\hbox{$\erg\s^{-1}\,$}}

\newcommand{\kmps}{\hbox{$\km\s^{-1}\,$}}

\newcommand{\halpha}{H$\alpha$}

\newcommand{\K}{{\rm K}}

\newcommand{\Mdot}{\hbox{$\dot M$}}
\newcommand{\Edot}{\hbox{$\dot E$}}

\newcommand{\pc}{{\rm\thinspace pc}}

\newcommand{\keV}{{\rm\thinspace keV}}

\newcommand{\Lsol}{\hbox{$\thinspace L_{\sun}$}}
\newcommand{\Msol}{\hbox{$\thinspace M_{\sun}$}}
\newcommand{\Zsol}{\hbox{$\thinspace Z_{\sun}$}}

\begin{document}

\title{Supernova feedback efficiency and mass loading in the starburst
and galactic superwind exemplar M82}

\author{David K. Strickland\altaffilmark{1} and
        Timothy M. Heckman\altaffilmark{1}
}

\altaffiltext{1}{Department of Physics and Astronomy,
        The Johns Hopkins University,
        3400 N.~Charles St., Baltimore, MD 21218, USA.}

\begin{abstract}
We measure the net energy efficiency of supernova and stellar wind feedback 
in the starburst galaxy M82 and the degree of mass-loading of 
the hot gas piston driving
its superwind by comparing a large suite of 1 and 2-dimensional hydrodynamical 
models to a set of observational constraints derived from hard X-ray 
observations of the starburst region (the fluxes of the He$\alpha$ and 
Ly$\alpha$-like lines of S, Ar, Ca and Fe, along with the total diffuse
$E=2$ -- 8 keV X-ray luminosity). 
These are the first direct measurements of the 
feedback efficiency and mass-loading of supernova 
heated and enriched plasma in 
a starburst galaxy. We consider a broad range of plausible parameters for 
the M82 starburst, varying the age and mode of star formation, the starburst 
region size and geometry, and supernova  metal yields. Over all these varied 
input parameters all the models that satisfy the existing observational 
constraints have medium to high thermalization efficiencies ($30$\% 
$\le \epsilon \le 100$\%) and the volume-filling 
wind fluid that flows out of the starburst region 
is only mildly centrally mass loaded
($1.0 \le \beta \le 2.8$). These results imply
a temperature of the plasma within the starburst region 
in the range 30 -- 80 million Kelvin, a mass flow rate of the wind
fluid out of the starburst region of
$\Mdot_{\rm tot} \sim 1.4$ -- $3.6 \Msol \pyr$ 
and a terminal
velocity of the wind in the range $v_{\infty} = 1410$ -- $2240 \kmps$. This 
velocity is considerably larger than the escape velocity from M82 
($v_{\rm esc} \la 460 \kmps$) and the velocity of the \halpha~emitting 
clumps and filaments within M82's wind ($v_{\rm H\alpha} \sim 600 \kmps$). 
Drawing on these results we provide a 
prescription  for implementing starburst-driven superwinds in cosmological 
models of galaxy formation and evolution that more accurately represents 
the energetics of the hot metal-enriched phases than the existing recipes 
do.
\end{abstract}



\section{Introduction}
\label{sec:introduction}

The presence of metals in the intergalactic medium (IGM) at all red-shifts
\citep{songaila97}, the present-day galaxy mass-metallicity relationship 
\citep{garnett02,tremonti04}, and $\sim 100$-kpc-scale holes in the IGM
at red-shift $z \sim 3$ \citep{adelberger03} 
have all been interpreted as the consequence
of energetic metal-bearing outflows from galaxies.

There is no doubt that at least one form of galactic wind exists:
starburst galaxies are
commonly observed to host supernova-driven galactic-scale winds
\citep{ham90,heckman03}
in both the local and high redshift Universe 
\citep{lehnert96a,pettini01,shapley03}. These superwinds are
phenomenologically 
complex and are comprised of multiple different gas phases moving at 
different velocities 
\citep{blandhawthorn95,dahlem97,heckman2000,heckman01,veilleux05}.
However, the very hot plasma long hypothesized to be the prime mover
in superwinds \citep{chevclegg}, often called the wind fluid, 
has only recently been directly observed
in the archetypal superwind associated with the starburst galaxy M82  
\citep{griffiths2000,strickland07}.

The most fundamental long-standing uncertainty in our understanding of the
nature of superwinds and their larger scale role is 
the value of the energy per particle of the metal-enriched 
wind-fluid that drives an individual superwind. 
Both supernovae (SNe) and stellar winds from massive stars are responsible
for supplying the kinetic energy that drive superwinds (Stellar winds 
only contribute $\sim 10$\% of the total
mechanical power of a starburst, but they may supply
up to half of the gas mass returned to the ISM, see 
\citealt{lrd92,leitherer_heckman95}).  What
fraction of the $\sim 10^{51} \erg$ of kinetic energy released 
per star that ultimately goes supernova 
is actually available to drive bulk motions of the ISM?
As it is understood that SNe dominate the energy return this is often
termed the SN thermalization efficiency or the SN feedback efficiency,
even though stellar winds do play some small part energetically.
Is the metal-enriched mix of SN and stellar wind ejecta diluted and cooled
by mixing with the ambient ISM, \ie is the wind-fluid mass-loaded? 
If this ejecta is to escape the galaxy and enter the IGM then 
its specific energy must exceed its initial gravitational 
potential energy, and furthermore the total energy in the wind-fluid must be 
sufficient to perform the mechanical work necessary to move the interstellar
and halo gas that lies between the starburst region and the IGM
\citep{maclow88,strickland04b}.

For moderate to high thermalization efficiency (efficiencies $\ga$ 10\%),
the wind-fluid is expected to have an initial temperature within
the starburst region in the range 10 -- 100 million Kelvin \citep{chevclegg},
even if it has been lightly mass-loaded with cold ambient gas.
Such temperatures are far hotter than the neutral, warm ionized or 
soft X-ray emitting plasmas that are the most easily 
observed phases within superwinds \citep{dahlem97,veilleux05}. 

However it has only recently become technically possible to detect the
diffuse hard X-ray emission that would be associated with a plasma 
in the temperature range $10^{7} \la \log T$ (K) $\la 10^{8}$. 
The starburst regions of even the nearest starburst galaxies
only subtend $\sim 15$ -- $45\arcsec$ on the sky and are heavily
populated with X-ray-luminous compact objects. It was only
with the advent of $\sim 1\arcsec$-spatial resolution X-ray astronomy
in 1999 following the launch of the {\it Chandra} X-ray Observatory 
that made it possible for \citet{griffiths2000} to resolve out the
point source emission in the archetypal nearby starburst galaxy M82
and reveal residual diffuse hard X-ray emission, although of uncertain origin.

It is important to 
distinguish between the material that actually drives a superwind, 
and the other more easily observed phases of swept-up
and entrained gas that are part of the wind. We define the wind-fluid
as the material whose
high thermal and/or ram pressure drives the superwind.
The wind fluid certainly comprises a significant fraction of the 
merged SN ejecta and massive star stellar wind material, 
probably admixed (mass-loaded\footnote{We will use the term mass loading to
denote the addition and efficient mixing of ambient ISM into the
full volume of the wind-fluid, which is consistent with the terminology
in \citet{suchkov96}. Technically mass loading need not affect the entire
flow and can be localized, see \eg \citet{massloadedflows5}. Furthermore
the form of mass-loading considered in this work occurs only within the
starburst region itself (``centralized mass loading'') although ultimately 
it affects the entire flow. We find it conceptually beneficial to 
differentiate between \emph{mass loading} and \emph{entrainment}
in superwinds. Mass loading affects the density of one gaseous phase
by mixing in material from another phase, almost always a cooler and
denser phase. The origin and fate of the secondary phase is unimportant
given the focus on the properties of the primary phase.
Entrainment is the incorporation into the superwind of subsidiary gas
phases that leads to creating or sustaining a multiphase superwind.
The entrained material is usually ambient disk or halo gas over-run
or swept-up by the wind fluid, and retains much of its original
character, albeit with some modification. Although entrainment processes
may lead to mass loading of the wind fluid the two processes are not
synonymous. In principle entrainment could occur without leading to
significant mass loading of the wind fluid. One can not realistically
hope to represent
or approximate the totality of a multiphase superwind as a single phase
mass loaded flow. Only specific components of the superwind are amenable
to such treatment, of which the wind fluid is the clearest example.})  
with some ambient gas from with
starburst region. 

Although other gas phases may dominate the total
gas mass budget of a superwind,
they are less important with respect
to the issue of energy and metal transfer into the IGM.
Furthermore the ejection of these phases is less than
certain, as the measured warm neutral and warm ionized medium (WNM, WIM)
gas velocities are typically comparable to the galactic escape velocities 
\citep{heckman2000,martin05,rupke05}. 
Unfortunately even the most recent attempts to 
include superwinds in cosmological models of galaxy formation
use only simplified single-phase winds, often 
with energetics and masses based on the
observed WNM/WIM properties of superwinds 
\citep[see \eg][]{springel03,bertone05}. These models
do not properly address the physics of hot-phase metal transport in addition to
having overly low values of the energy per particle, which casts the
reliability of their conclusions into doubt.

Attempts have been made to use
measurements of the temperature of the extended 
soft X-ray emitting plasmas ($T \sim 2$ -- $7 \times 10^{6}$ K)
in starburst galaxies to constrain whether the hot gas will 
escape the host galaxy \citep[see \eg][]{martin99}, but such an
approach neglects both the significant kinetic energy expected to
be associated with this material and 
that the soft X-ray emitting
plasma is an indirect probe of the properties of the
true wind-fluid \citep{ss2000}.

The detection of apparently thermal diffuse hard X-ray emission of
temperature $\sim 4 \times 10^{7}$ K in the starburst region of M82 
in the 33 ks-long {\it Chandra} ACIS-I observations of \citet{griffiths2000}
suggested it was possible to directly observe and constrain the 
properties of the wind fluid in a starburst superwind.

We recently obtained and analyzed a new 18 ks {\it Chandra} ACIS-S 
observation
of M82, and also reprocessed archival {\it Chandra} ACIS-I and 
{\it XMM-Newton} observations\footnote{{\it Chandra} ObsId numbers 
\dataset[ADS/Sa.CXO#Obs/2933]{2933},
\dataset[ADS/Sa.CXO#Obs/361]{361} and \dataset[ADS/Sa.CXO#Obs/1302]{1302}.
The {\it XMM-Newton} data ObsIds are 0112290201 and 0206080101.}
 (total exposure of 48 ks with ACIS-I, 75 ks with
{\it XMM-Newton} PN and 102 ks with each MOS instrument).
We confirmed the presence of 
both diffuse continuum and diffuse $E\sim 6.7 \keV$ Fe He$\alpha$
line emission within the central starburst region 
\citep[hence forth referred to as Paper I]{strickland07}. Having assumed 
that the line-emitting plasma is metal-enriched to 
$Z_{\rm Fe} \sim 5 Z_{\rm Fe, \odot}$, we found that the
observed iron line
luminosity of $L_{X, 6.7} \sim (1.1$ -- $1.7) \times 10^{38} \ergps$
was consistent with the properties of the 
wind fluid expected given M82's star formation
rate, provided that the SN thermalization efficiency was high. 
However the diffuse 
hard X-ray continuum luminosity of $L_{\rm X, 2-8} \sim 4.4 \times 10^{39}
\ergps$ was too luminous to be the thermal bremsstrahlung associated with
the wind fluid given the parameters assumed when interpreting the 
iron line emission.

In this paper we consider in greater detail what constraints the
observed hard X-ray properties of M82 can place on the properties
on the superwind and the wind-fluid, in particular the
efficiency of SN heating ($\epsilon$) and the degree of central mass-loading
($\beta$). Ultimately these determine the energy
per particle of the wind fluid. We restrict analysis to the starburst
region within $r \sim 500$ pc of the center of M82. We calculate the
X-ray emission using both multi-dimensional hydrodynamical simulations
and the 1-dimensional analytical \citet{chevclegg} model
(hence forth referred
to as the CC model).
By using the simple but relatively accurate CC model we can 
explore the predicted X-ray properties of the wind fluid in models covering 
a wider range of parameter space 
than if we used multi-dimensional hydrodynamical simulations alone.

A variety of recent papers have presented theoretical
calculations of the broad-band X-ray emission from
individual star-forming clusters, using either the adiabatic
CC model, or the non-adiabatic solution of Silich and collaborators
\citep{canto00,stevenshartwell03,oskinova05,silich05,ji06}, with
the aim of seeing if the particular model used can explain the observed data.
Our motivation differs from these papers, in that our aim is to use
theoretical models  in combination with observational data on M82 to 
constrain fundamental properties of the superwind. In addition
our approach differs from these previous papers in its use of X-ray
line luminosities rather than purely broad-band X-ray luminosities.
In reality the broad-band emission might be dominated by 
non-thermal continuum sources. Furthermore line emission is
more sensitive than the thermal continuum luminosity to 
interesting parameters such as the plasma temperature.

\section{Observational constraints on high temperature gas in M82}
\label{sec:observations}

The observational constraints derived in Paper I on the
diffuse hard X-ray emission within $r = 500$ pc of the nucleus
of M82 from the {\it Chandra} and {\it XMM-Newton} observations 
are summarized in Table~\ref{tab:paper1}. These comprise a robust 6.69
keV iron line detection ($> 4\sigma$ in significance), 
an upper limit on the 6.96 keV
iron line, and from the {\it Chandra} observations
 the luminosity of the diffuse
continuum. For the {\it Chandra}-based 
observations the values refer to the diffuse emission alone,
after the exclusion of detected point-like X-ray sources and
corrections for undetected point sources based on the observed
point source luminosity function.
These luminosities also
include 32\% and 20\% upward corrections to the luminosities 
derived from the ACIS-S and ACIS-I data to account for 
regions excluded from the spectra due to remove the emission
from detected point sources.
For the {\it XMM-Newton} observation the line luminosities include
all emission within $r = 500$ pc, but the contribution to
the line luminosities from
point-like sources is expected to be small in this particular observation.

\begin{deluxetable*}{lrrrrr}
 \tabletypesize{\scriptsize}%
\tablecolumns{6}
\tablewidth{0pc}
\tablecaption{Diffuse continuum and iron line luminosities from Paper 1.
        \label{tab:paper1}}
\tablehead{
\colhead{Instrument} & \colhead{ObsID}
     & \colhead{$\log L_{\rm 2-8 keV}$}
     & \colhead{$\log L_{\rm Fe K\alpha}$} 
     & \colhead{$\log L_{\rm Fe He\alpha}$} 
     & \colhead{$\log L_{\rm Fe Ly\alpha}$}
     \\
\colhead{(1)}
     & \colhead{(2)} & \colhead{(3)}
     & \colhead{(4)} & \colhead{(5)} & \colhead{(6)}
}
\startdata
{\it XMM-Newton} EPIC & 0206080101
     & \nodata
     & $37.31_{-0.54}^{+0.23}$
     & $38.13_{-0.06}^{+0.05}$
     & $<37.66$ \\
{\it Chandra} ACIS-S & 2933
     & $39.65_{-0.05}^{+0.02}$
     & $37.94_{-0.48}^{+0.22}$
     & $38.23_{-0.27}^{+0.16}$
     & $<38.15$ \\
{\it Chandra} ACIS-I & 361+1302
     & $39.64_{-0.02}^{+0.02}$
     & $37.93_{-0.26}^{+0.20}$
     & $38.13_{-0.25}^{+0.14}$
     & $<38.01$ \\
\enddata
\tablecomments{This table summarizes the results presented in 
 \citet{strickland07} on the emission from within $r = 500$ pc
  of the nucleus of M82.
 Column 2: The ObsID is the identification number allocated
 to each observation by the {\it Chandra} and {\it XMM-Newton}
 Science Centers.
 Column 3: Logarithm of the total diffuse 
 X-ray luminosity in the $E=2$ -- 8 keV
 energy band, including line emission.
 Column 4: Logarithm of the $E\sim 6.4$ keV Fe K$\alpha$ line luminosity. 
 Column 5: Logarithm of the $E\sim 6.7$ keV Fe He$\alpha$ line luminosity. 
 Column 6: Logarithm of the $E\sim 6.96$ keV Fe Ly$\alpha$ line luminosity. 
 The quoted uncertainties are 68.3\% confidence for one interesting 
 parameter, while the upper limits are 99.0\% confidence.
}
\end{deluxetable*}

\begin{figure}[!t]
\plotone{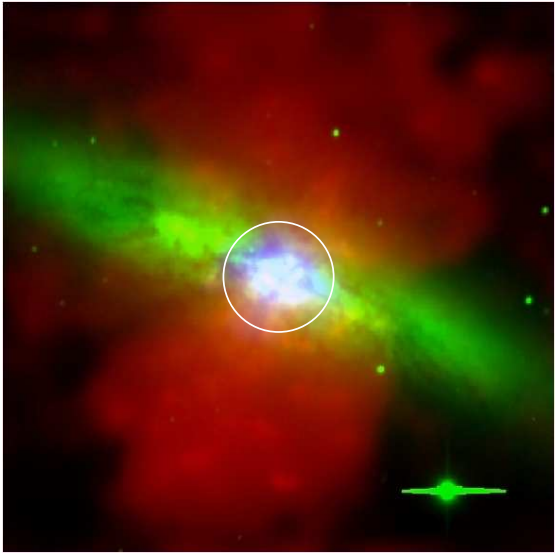}
  \caption{A three color composite image of M82 showing the region within
a 500 pc radius of the nucleus (white circle) in comparison to the galaxy
and superwind. Soft X-ray emission in the 0.3--2.8 keV energy band is shown in
red, optical R-band emission (starlight) 
in green, and diffuse hard X-ray emission in the
3 -- 7 keV energy band in blue. The X-ray images are adaptively smoothed
Chandra ACIS-S images with point source emission removed and interpolated over.
The image is $4\farcm8$ ($\sim 5$ kpc) on a side.
  }
  \label{fig:m82_3color}
\end{figure}

Note that the region of bright diffuse hard X-ray emission studied
in this paper is associated with the nuclear starburst region. 
As can be seen from Fig.~\ref{fig:m82_3color} this 
region is much smaller than the full 12-kpc-scale superwind that
is most often studied in optical \citep{lehnert99} or soft X-ray emission 
\citep{stevens03}.

As discussed in Paper I the nature of the diffuse broad-band continuum
is not clear. 
If the continuum is thermal then the relative strength
of the iron line emission to the observed continuum is much lower than 
expected for a plasma of Solar elemental abundances, let alone the 
iron-enriched plasma expected from recent core-collapse SNe. We
demonstrated that a wind mass-loaded with additional ambient gas
could produce the total broad-band
X-ray luminosity that we observe, but the required degree of mass loading
could not sufficiently reduce the relative strength of the iron line
to the continuum.
Thus the best fit bremsstrahlung models for the continuum do not
necessarily provide meaningful estimates of the temperature of the
iron-emitting plasma, and should be interpreted with caution.
Power law spectral models provide a marginally better
fit to the continuum than do thermal bremsstrahlung models. 
Thus a non-thermal process such as Inverse Compton X-ray emission might be 
responsible for the majority of the diffuse hard X-ray continuum.
The X-ray luminosity associated with Inverse Compton emission 
depends sensitively on the unknown magnetic field strength within the starburst
region. In Paper I we estimated it might
account for $\sim 25$\% of the observed diffuse  X-ray 
luminosity in the $E=2$ -- 8 keV energy band, although other
authors have in the past given higher luminosity estimates for Inverse Compton 
emission from M82 \citep{schaaf89,moran97}.

The upper limits on the iron 6.96 keV / 6.69 keV line flux 
ratio\footnote{Note that the line ratios quoted in Paper I were the ratio of
the photon fluxes, whereas in this paper we quote the energy flux ratio.} of 
$< 0.34$ place upper limits on the temperature of $T < 7.8 \times 10^{7} \K$ 
if the plasma is in collisional equilibrium 
(see Table~\ref{tab:temp_estimates}). The Coulomb relaxation
and ionization equilibrium timescales presented in Paper I are much
shorter than the timescale for plasma to flow out of the starburst region,
so the assumption of collisional ionization equilibrium is reasonable.

\begin{deluxetable*}{lll}
 \tabletypesize{\scriptsize}%
\tablecolumns{6}
\tablewidth{0pc}
\tablecaption{Temperature estimates for the diffuse hard X-ray emission.
        \label{tab:temp_estimates}}
\tablehead{
\colhead{Basis for temperature estimate} & \colhead{Detector}
     & \colhead{$\log T$ (K)}
     \\
\colhead{(1)}
     & \colhead{(2)} & \colhead{(3)}
}
\startdata
Fe Ly$\alpha$ / Fe He$\alpha$ & XMM-Newton & $<7.89$ \\
Fe Ly$\alpha$ / Fe He$\alpha$ & ACIS-S     & $<8.05$ \\
Fe Ly$\alpha$ / Fe He$\alpha$ & ACIS-I     & $<8.04$ \\
Ca He$\alpha$ / Fe He$\alpha$ & XMM-Newton & $7.31_{-0.06}^{+0.10}$ \\
Ca He$\alpha$ / Fe He$\alpha$ & ACIS-S     & $>7.21$ \\
Ar He$\alpha$ / Fe He$\alpha$ & XMM-Newton & $7.30\pm{0.03}$ \\
Ar He$\alpha$ / Fe He$\alpha$ & ACIS-S     & $>7.29$ \\
S He$\alpha$ / Fe He$\alpha$  & XMM-Newton & $7.26\pm{0.01}$ \\
S He$\alpha$ / Fe He$\alpha$  & ACIS-S     & $7.23_{-0.03}^{+0.06}$ \\
(Ca Ly$\alpha$ + Ca He$\alpha$) / Fe He$\alpha$ & XMM-Newton & $>7.22$ \\
(Ca Ly$\alpha$ + Ca He$\alpha$) / Fe He$\alpha$ & ACIS-S     & $>7.12$ \\
(Ar Ly$\alpha$ + Ar He$\alpha$) / Fe He$\alpha$ & XMM-Newton & $>7.26$ \\
(Ar Ly$\alpha$ + Ar He$\alpha$) / Fe He$\alpha$ & ACIS-S     & $>7.20$ \\
(S Ly$\alpha$ + S He$\alpha$) / Fe He$\alpha$ & XMM-Newton 
     & $7.30_{-0.02}^{+0.01}$ \\
(S Ly$\alpha$ + S He$\alpha$) / Fe He$\alpha$ & ACIS-S     & $>7.26$ \\
S He$\alpha$ / Ca He$\alpha$  & XMM-Newton & $7.22_{-0.05}^{+0.07}$ \\
S He$\alpha$ / Ca He$\alpha$  & ACIS-S     & $<7.27$ \\
S He$\alpha$ / Ar He$\alpha$  & XMM-Newton & $7.19\pm{0.05}$ \\
S He$\alpha$ / Ar He$\alpha$  & ACIS-S     & $<7.11$ \\
Ar He$\alpha$ / Ca He$\alpha$ & XMM-Newton & $7.28_{-0.16}^{+0.13}$ \\
Ca Ly$\alpha$ / Ca He$\alpha$ & XMM-Newton & $<7.63$ \\
Ca Ly$\alpha$ / Ca He$\alpha$ & ACIS-S     & $<7.69$ \\
Ar Ly$\alpha$ / Ar He$\alpha$ & XMM-Newton & $<7.42$ \\
Ar Ly$\alpha$ / Ar He$\alpha$ & ACIS-S     & $<7.59$ \\
S Ly$\alpha$ / S He$\alpha$   & XMM-Newton & $7.10_{-0.04}^{+0.02}$ \\
S Ly$\alpha$ / S He$\alpha$   & ACIS-S     & $<7.12$ \\
Bremsstrahlung fit to diffuse continuum & ACIS-S     & $7.64_{-0.16}^{+0.22}$ \\
Bremsstrahlung fit to diffuse continuum & ACIS-I     & $7.50_{-0.07}^{+0.09}$ \\
\enddata
\tablecomments{
 Temperature estimates are given only for 
 line ratios where one or more of the lines involved is detected at
 $\ge 3\sigma$.
 The quoted uncertainties are 68.3\% confidence for one interesting 
 parameter, while both upper and lower limits are 99.0\% confidence.
}
\end{deluxetable*}

Other lines in the $2 \le E$ (keV) $\le 10$ hard X-ray band may in 
principle be used to constrain the plasma temperature in a starburst
region. In particular the helium-like and hydrogen-like
ions of sulphur (at $E \sim 2.45$ keV and $E \sim 2.62$ keV respectively), 
argon ($E \sim 3.13$ keV and $E \sim 3.33$ keV) and calcium ($E \sim 3.90$ 
keV and $E \sim 4.11$ keV) can be detected in
CCD spectra from the {\it Chandra} ACIS and {\it XMM-Newton}
EPIC instruments. These lines are most intense in plasmas with temperatures
in the range $\log T \sim 7.0$ -- 7.5, somewhat lower than the optimal
temperature for the iron lines, but this might be advantageous in
diagnosing starbursts where mass loading is significant and/or the efficiency
of SN thermalization is low. 

\begin{figure*}[!t]
\plotone{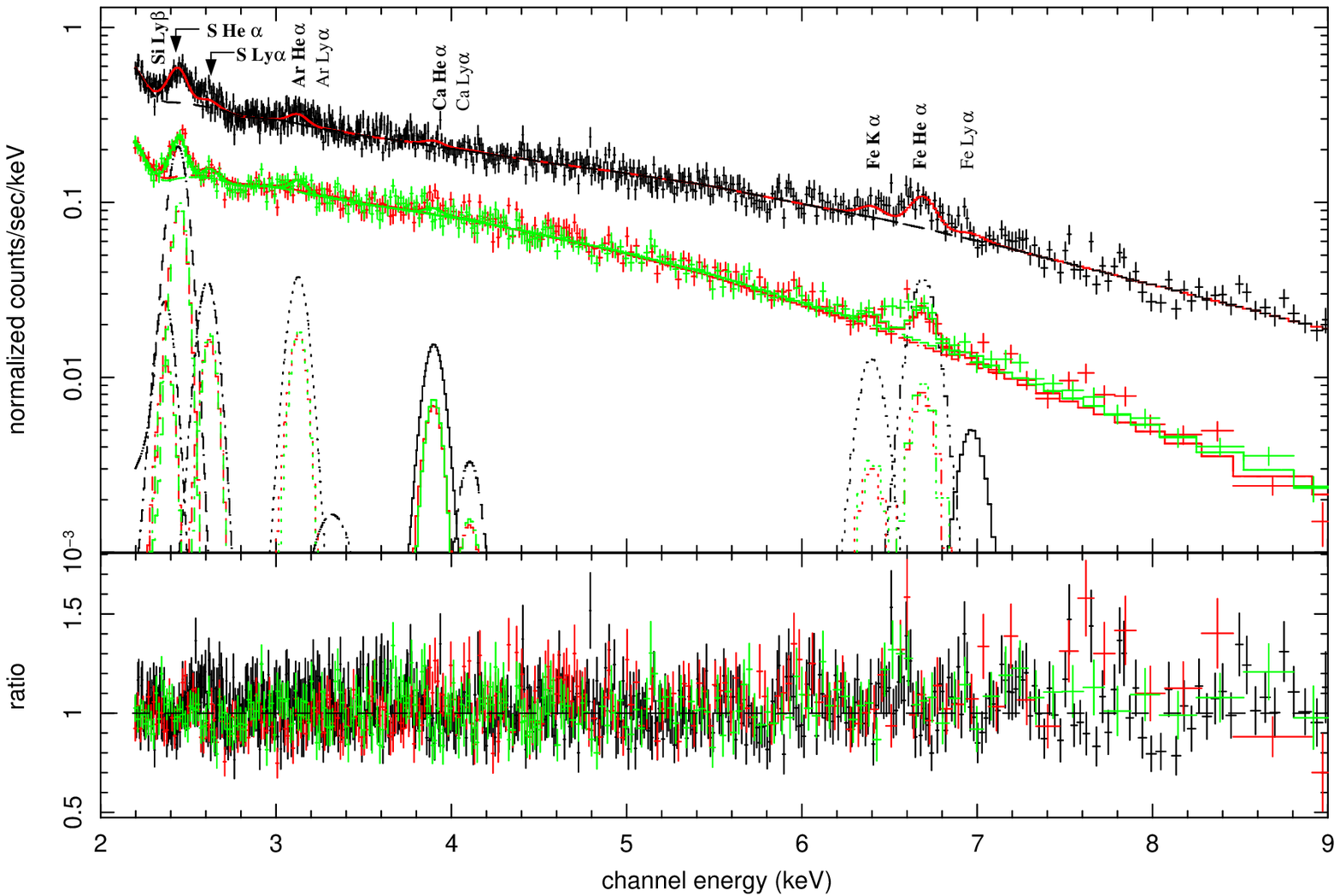}
  \caption{(Upper panel) The {\it XMM-Newton}
  EPIC pn (black data points with error
  bars), MOS1 (red data points) and MOS2 (green data points) spectra
  of the central 500 pc of M82. The best-fit spectral model (red solid line)
  is shown, along with the named X-ray line components of the model 
  (dashed and dotted lines with a color matching the parent data).
  The He$\alpha$-equivalent lines of S, Ar, Ca and Fe, along with 
  Si Ly$\beta$ and S Ly$\alpha$, are all detected
  with significances of $\ga 3\sigma$. The other lines have lower significance.
  (Lower panel) The residuals from the best-fit model.
  }
  \label{fig:xmm_spectrum}
\end{figure*}

In Paper I we specifically excluded data at the expected energies
 of the S, Ar and
Ca lines from the fitting of the M82 nuclear region spectra. 
Repeating the analysis described
Paper I but now fitting all the data in the energy range $E=2.2$ 
-- 9.0 keV in both the {\it XMM-Newton} and {\it Chandra} ACIS-S spectra 
yields the line detection significances and line luminosities
given in Table~\ref{tab:s_ar_ca}. We fit for the intensity of the helium-like
and hydrogen-like S, Ar and Ca lines, having fixed the line energies at
the expected values. These lines are faint, but in some cases likely to be
real. As an example we display the best-fit to the {\it XMM-Newton} EPIC 
pn + MOS spectra
in Fig.~\ref{fig:xmm_spectrum}.
Detection significance was assessed using Monte Carlo
methods with several thousand simulated spectra. 
Where the line detection significance was $< 3 \sigma$ we give
the 99\% upper limit on the line luminosity (and not the best-fit value).
The {\it Chandra} ACIS-I spectra gave similar results to those from ACIS-S,
but as the line detections are less significant than in the {\it XMM-Newton}
data there is little reason to present the values.

\begin{deluxetable*}{llrlll}
 \tabletypesize{\scriptsize}%
\tablecolumns{6}
\tablewidth{0pc}
\tablecaption{Hard X-ray S, Ar and Ca line detections and luminosities in M82.
        \label{tab:s_ar_ca}}
\tablehead{
\colhead{Line} & \colhead{Ion} & \colhead{Energy}
     & \colhead{Detector}
     & \colhead{Significance} & \colhead{$\log L_{\rm X}$}
     \\
\colhead{(1)}
     & \colhead{(2)} & \colhead{(3)}
     & \colhead{(4)} & \colhead{(5)}
     & \colhead{(6)}
}
\startdata
He$\alpha$ & S XV & 2.45
  & XMM-Newton & $>3.5$ & $38.19\pm{0.02}$ \\
S Ly$\alpha$ & S XVI & 2.62
  & XMM-Newton & $>3.5$ & $37.43_{-0.10}^{+0.08}$ \\
Ar He$\alpha$ & Ar XVII & 3.13
  & XMM-Newton & $>3.5$ & $37.53_{-0.08}^{+0.07}$ \\
Ar Ly$\alpha$ & Ar XVIII & 3.32
  & XMM-Newton & $0.8$ & $<37.26$ \\
Ca He$\alpha$ & Ca XIX & 3.90
  & XMM-Newton & $3.5$ & $37.27_{-0.17}^{+0.12}$ \\
Ca Ly$\alpha$ & Ca XX & 4.11
  & XMM-Newton & $1.1$ & $<37.14$ \\
S He$\alpha$ & S XV & 2.45
  & ACIS-S & $>3.9$ & $38.32_{-0.07}^{+0.06}$ \\
S Ly$\alpha$ & S XVI & 2.62
  & ACIS-S & $1.7$ & $<37.64$ \\
Ar He$\alpha$ & Ar XVII & 3.13
  & ACIS-S & $1.9$ & $<37.56$ \\
Ar Ly$\alpha$ & Ar XVIII & 3.32
  & ACIS-S & $0.5$ & $<37.69$ \\
Ca He$\alpha$ & Ca XIX & 3.90
  & ACIS-S & $2.4$ & $<37.52$ \\
Ca Ly$\alpha$ & Ca XX & 4.11
  & ACIS-S & $0.5$ & $<37.54$ \\
\enddata
\tablecomments{ 
The luminosities are given as $\log_{10}$ of the line
  luminosity in units of $\ergps$, assuming a distance to M82
  of $D=3.6$ Mpc and zero intrinsic absorption. Line energies are
in units of keV. Line significances are the $\sigma$ values equivalent
  to the enclosed probability $1-p$ for a two-tailed Gaussian, where
  $p$ is the probability of the line being spurious. In the case of highly
  significant lines we can only estimate a lower limit to the significance
  of the line.
  The quoted uncertainties on the luminosities are 68.3\% confidence, 
  while upper limits on $\log L_{\rm X}$ are provided 
  at 99.0\% confidence if the line significance is $<3\sigma$.
}
\end{deluxetable*}

In the {\it XMM-Newton} EPIC pn + MOS 
spectra of M82's nucleus the equivalent of
the He$\alpha$ line from highly ionized S, Ar, Ca and Fe are all detected
with significances $\ga 3\sigma$, while in the shorter (and hence lower S/N)
{\it Chandra} ACIS-S spectrum of the nuclear diffuse emission the S 
He$\alpha$ and Fe He$\alpha$ are detected at $\ga 3\sigma$ while the
Ar He$\alpha$ and Ca He$\alpha$ lines are only detected at 
$\la 2\sigma$. In both instruments Si Ly$\beta$ is also detected at
high significance, but we shall ignore this 
lower ionization state line in the analysis to follow. The only 
Ly$\alpha$-like line detected at $\ga 3\sigma$ is S Ly$\alpha$ in the
{\it XMM-Newton} EPIC spectra.

Using the robust line detections (\ie those with significance $\ge 3 \sigma$)
we can estimate the plasma temperature from
the ratio of S Ly$\alpha$ to S He$\alpha$, and from between different
elements from the ratio of the various He$\alpha$ lines (assuming
their relative elemental abundances are Solar). These temperature estimates
are presented in Table~\ref{tab:temp_estimates}. The S ions tightly
constrain the ion temperature to $\log T \approx 7.1$, while the ratio
of the S, Ar or Ca He$\alpha$ line to the Fe He$\alpha$ line yields
$\log T \sim 7.3$. The ratio of S He$\alpha$ to either Ar He$\alpha$
or Ca He$\alpha$ yields a temperature estimate of $\log T \sim 7.2$.

This pattern of the inferred temperature increasing as higher ionization
state ions are used (which are of course 
more sensitive to higher temperatures) 
indicates that either the plasma
is multi-phase, and/or the the plasma is not in collisional ionization
equilibrium (CIE). Of these explanations we favor the former, as 
multi-dimensional numerical hydrodynamical simulations predict a broad
range of gas temperatures below $\log T \sim 8$ \citep[see \eg][]{ss2000}. 
Furthermore we discussed in Paper I reasons to expect the Fe-emitting
plasma to be in or close to CIE. 

Unfortunately these results do not strongly constrain the temperature of
the Fe-emitting plasma. One possibility is that the iron line emission
arises in the very hottest phases in the superwind, 
\ie at $\log T \la 7.9$, and that
most of the S, Ar and Ca emission comes from somewhat cooler components
associated with the wind/ISM interaction and the larger scale soft
X-ray emission in the M82 superwind. In this case, given the possible
soft X-ray contribution, the S, Ar and Ca observed
luminosities must be treated as upper limits to the S, Ar and Ca line emission
coming from the volume filling wind-fluid.
Alternatively the wind fluid
is relatively cool, with a temperature corresponding to that
measured from the Ar and Ca lines, \ie $\log T \sim 7.3$, and the Fe lines
are produced by some unexplained non-thermal mechanism.

In either
case a successful model of M82's nuclear emission must reproduce
the observed hard X-ray iron line emission without over-producing the
broad-band continuum and S, Ar and Ca line emission.

\section{Theoretical and Numerical Methodology}
\label{sec:theory}

Hard X-ray emission in multi-dimensional hydrodynamical models
of superwinds is dominated by the thermal emission from 
starburst region itself \citep{tb93,suchkov94,ss2000}. 

If the starburst region is spherical then the plasma
properties of this region obtained from
hydrodynamical simulations agree very well
with the simple one-dimensional radially-symmetric 
analytical CC model, in which the mechanical energy and returned 
gaseous ejecta from core-collapse supernovae 
are injected at a uniform rate only within a radius $R_{\star}$,
and radiative energy losses are assumed to be negligible.
Gravity and any surrounding ambient gaseous medium are ignored.

As the feedback parameters we wish to constrain are related to the observable
 X-ray emission via the plasma properties in the starburst region it
is worthwhile for us to explicitly reproduce the analytical solution 
obtained by \citet{chevclegg}.
We consider the conditions under which
the adiabatic assumption inherent in the CC
model may be invalid in Appendix~\ref{app:radiative_case}, 
but we will show that for most
reasonable parameters associated with the M82 starburst (described below) the 
radiative energy losses are not significant.

In real galaxies starburst regions are better described as disks or ring
of star-formation rather than spherical regions. To explore the predicted
X-ray emission for our default model of the M82's disk-like starburst we begin
by using 2-dimensional hydrodynamical simulations in cylindrical symmetry to
predict the plasma properties within the a radius of 500 pc of the center of
the galaxy.

However, with appropriate scaling (see Appendix~\ref{app:non_spherical}), 
it is possible to use the 1-dimensional
CC model with a spherical starburst to predict the plasma properties 
of the wind fluid within
a non-spherical starburst region to high accuracy, and to predict the 
hard X-ray emission from even larger regions with an accuracy no worse than the
uncertainties in the observed X-ray luminosities.

These scaled CC models are considerably faster to calculate than a full
hydrodynamical model, so we will use them to explore how the constraints on
the thermalization efficiency and mass loading change if we adopt 
a different range of parameters for the M82 starburst.

This approach, in particular the use of the scaled CC model, is applicable
only to the study of the wind fluid in the vicinity of the starburst
region. It is not intended for use as a model of an entire superwind
as it neglects the interaction between the wind fluid and the multiphase
ambient ISM.
The simplification of neglecting the ambient ISM works
because the only significant source of thermal hard X-ray emission
in a superwind is the wind fluid itself, in particular the wind fluid
within the starburst region. The interactions between the wind fluid and
the multiphase ISM are important in the large scale wind and for
its emission at soft X-ray and longer wavelengths 
\citep[\eg][]{ss2000,cooper08}, but with the exception of altering
the net efficiency of SN+SW heating and any mass-loading of the wind fluid
these interactions are not expected to
alter the properties of the wind fluid within the starburst region to
any significant degree given its energetic dominance.

\subsection{2-Dimensional Hydrodynamical Modeling}
\label{sec:hydromodel}

The hydrodynamical simulations were performed using the 
multi-dimensional numerical hydrodynamics code VH-1 \citep{blondin90}.
Simulations were performed in 2-dimensional cylindrical symmetry with a
cell size of 2 pc over 500 by 500 cell computational grid. Reflecting boundary
conditions were imposed along the $r$ and $z$ axes, with inflow/outflow
boundary conditions are the other edges of the computational grid.

Energy and mass are injected within the starburst region in the manner
described in \citet{ss2000}. In general the starburst region was modeled
as disk-like region of the size described in \S~\ref{sec:theory:applying}.
No gravitational forces or radiative cooling
are considered for the purposes of consistency with the analytical CC model
described below.

The computational grid was initially populated with a tenuous ISM of
density $\rho = 2 \times 10^{-28} \gpcc$ and temperature 
$T = 8 \times 10^{3}$ K. The simulations were run until the wind had
blown the initial ISM off the grid and reached a steady-state solution,
which typically occurred $t \sim 1$ -- 2  Myr after the onset of the
simulation. Gas properties and
X-ray luminosities are measured at a time of 3 Myr within a 
spherical region of radius of 500 pc to closely match the observed region
discussed in \S~\ref{sec:observations}.

\subsection{The 1-Dimensional Analytical CC Solution}
\label{sec:ccmodel}

For completeness we present the analytical flow solution derived by CC, 
and how we use it to derive the fluid properties as a function of radial
position. 
More complicated, but also more generalized, 
methods of deriving the solution are presented in \citet{canto00} 
and \citet{silich04}. The basic model is completely specified by
only three parameters: the energy and
mass injection rates ($\Edot_{\rm tot}, \Mdot_{\rm tot}$); 
and the radius of the starburst region ($R_{\star}$) within which the
energy and mass are injected. 

CC present the stationary 
solution to the hydrodynamical equations as a function
of the Mach number $M = v/c_{\rm s}$, where $v$ is the local speed of the
flow and the sound speed is $c_{\rm s} = \surd (\gamma P/\rho)$. The 
pressure $P$, density $\rho$ and adiabatic index $\gamma$ 
have their normal meanings. At radii $r < R_{\star}$ mass and 
energy are injected at a uniform constant rate per unit volume $q_{M} = 
\Mdot_{\rm tot}/V_{\star}$ and 
$q_{E} = \Edot_{\rm tot}/V_{\star}$, where the starburst volume
$V_{\star} = 4\pi R_{\star}^{3}/3$, and the solution for the Mach number is
\begin{equation}
  \left( \frac{3\gamma+1/M^{2}}{1+3\gamma} 
  \right)^{-\frac{(3\gamma+1)}{(5\gamma+1)}} \,
\left( \frac{\gamma - 1 + 2/M^{2}}{1+\gamma} 
\right)^{\frac{(\gamma+1)}{2(5\gamma+1)}} 
 = \frac{r}{R_{\star}}.
 \label{equ:cc_smallr}
\end{equation}

This approach to the energy and mass injection
implicitly assumes that the exact manner in
which stars return mechanical energy to the ISM can be ignored due to rapid
local thermalization and mixing, \ie individual stellar winds and SN blast 
waves collide and merge on spatial scales small compared to the size of
the starburst region. If radiative energy losses are to be small then these
collisions must occur with a time shorter than the cooling time scale
for stellar wind bubbles or supernova remnants (SNRs). Theoretical
investigations of some of these issues may be found in \citet{larson74},
\citet{ham90}, \citet{canto00}, \citet{melioli04} and \citet{melioli05}.

At larger radii $r > R_{\star}$ then $q_{M} = q_{E} = 0$, and
the solution is
\begin{equation}
 M^\frac{2}{\gamma-1} \,
\left( \frac{\gamma - 1 + 2/M^{2}}{1+\gamma} 
\right)^{\frac{(\gamma+1)}{2(\gamma-1)}} 
 = \left( \frac{r}{R_{\star}} \right)^{2}.
\label{equ:cc_larger}
\end{equation}

For any given fractional radius $r/R_{\star}$ the Mach number is
found using a numerical root-finding technique (in our case the 
Newton-Raphson method) on the appropriate equation.

To obtain the density at any radius $r < R_{\star}$ we integrate Equation~1
of CC to obtain
\begin{equation}
\rho = \frac{q_{M} \, r}{3 \, v}.
\label{equ:cc_rho_smallr}
\end{equation}

Within the starburst region 
the gas density, pressure and temperature vary only slowly with radius,
with maxima at $r = 0$. The central density is effectively determined by
the rate at which the injected material can flow out of the starburst
region, $\rho_{\rm c} \propto \Mdot_{\rm tot} / (4\pi 
R_{\star}^{2} \, v_{\infty})$ where $v_{\infty}$ is the terminal velocity
of the wind at large radius, and the central temperature is set by the 
energy injected per particle. More specifically,
the central temperature and density 
are $T_{\rm c} = 0.4 \, (\mu m_{\rm H} \Edot_{\rm tot}) / 
(k \Mdot_{\rm tot})$ and 
$\rho_{\rm c} = 0.296 \, \Mdot_{\rm tot}^{3/2} \Edot_{\rm tot}^{-1/2} 
R_{\star}^{-2}$ for $\gamma = 5/3$.

At radii $r \ge R_{\star}$ the mass flow rate through the spherical surface
of radius $r$ is equal to the total mass injection rate within the starburst
region, so the density is
\begin{equation}
\rho = \frac{q_{M} \, R_{\star}^{3}}{3 \, v \, r^{2}}.
\label{equ:cc_rho_larger}
\end{equation}

To derive the local velocity of the flow $v$ we must first determine 
the speed of sound. If we make the assumption that the flow is 
adiabatic then we can obtain
the speed of sound at any point ($r < R_{\star}$ or $r \ge R_{\star}$), 
as we know the Mach number and the initial energy per particle is conserved.
Specifically we solve Equation 3 of CC and substitute for $\rho$ using 
Equation~\ref{equ:cc_rho_smallr}, to obtain
\begin{equation}
c_{s}^{2} \, \left( \frac{M^{2}}{2} + \frac{1}{\gamma-1}\right) 
= \frac{\Edot_{\rm tot}}{\Mdot_{\rm tot}}.
\label{equ:cc_csound}
\end{equation}
The speed of the flow at any point is then simply $v = M c_{\rm s}$. 
For $r < R_{\star}$ the flow is subsonic, with $u \propto r$.  The sonic
point is reached at $r = R_{\star}$, beyond which the flow velocity rapidly
tends toward the terminal velocity
$v_{\infty} = \surd 2 \, \Edot_{\rm tot}^{1/2} \, \Mdot_{\rm tot}^{-1/2}$.

With the density $\rho$ and the speed of sound $c_{\rm s}$ determined
at every point within the flow the thermal pressure $P$ is also now known.
The kinetic temperature $T = \mu m_{\rm H} \, P / \rho \, k$, where
$\mu m_{H} \approx 1.02 \times 10^{-24} \gpcc$ is the mean mass per particle
for a fully ionized plasma of Solar composition.

Numerically we derive
the gas density, temperature and velocity as a function of radius,
iterating outward in thin radial shells. The numerical implementation we
use is based on code originally created by
Ian Stevens \citep{stevenshartwell03}. This semi-analytical
code reproduces the density and pressure to within 0.1\% of
the values predicted by the full 2-D hydrodynamical simulation using the VH-1
code for a spherical starburst region.
The scaling used to approximate
a non-spherical starburst region using the spherical CC model is described
in Appendix~\ref{app:non_spherical}. 

Figure~\ref{fig:radial_solution} presents the radial variation in  several
fluid variables and associated derived quantities for one particular set
of input parameters, as predicted using our 1-dimensional CC model code.

\begin{figure*}[!t]
\plotone{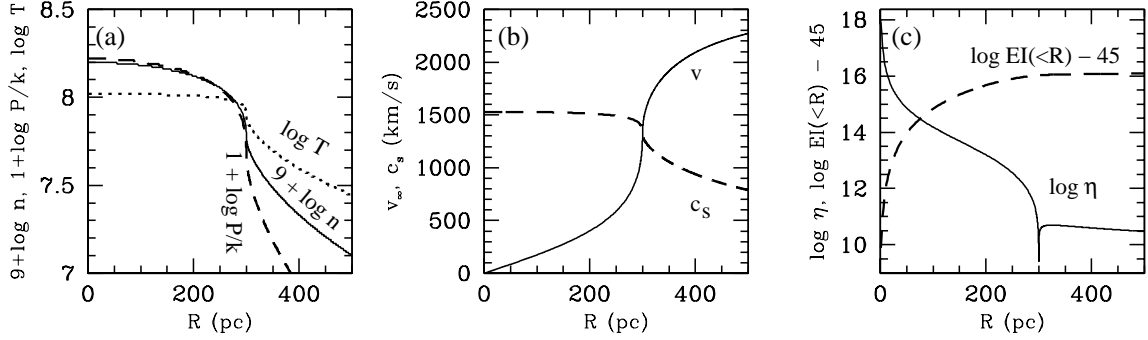}
  \caption{(a) An example of the radial CC solution described in 
  \S~\ref{sec:ccmodel}. Note the relatively uniform number
  density, thermal pressure and temperature within the assumed starburst
  region of radius $R_{\star} = 300$ pc, and rapid drop in those 
  parameters outside the starburst region. The net energy and mass injection
  rates assumed are $\Edot_{\rm tot} = 3.1 \times 10^{42} \ergps$ and
  $\Mdot_{\rm tot} = 1.4 \Msol \pyr$.
  (b) Local wind velocity $v$ and sound speed
  $c_{\rm s}$.
  (c) Logarithms of the ionization-state-related parameter $\eta$ 
  (Equation~\ref{equ:eta}) and cumulative volume emission integral $EI(<R)$.
  The plasma is  in ionization 
  equilibrium when $\log \eta \ge 12$. Values are given in c.g.s units unless
  it is stated otherwise.}
  \label{fig:radial_solution}
\end{figure*}

\subsection{Parameters of the M82 nuclear starburst}
\label{sec:theory:applying}

\subsubsection{The starburst age and magnitude}
\label{sec:theory:applying:sbmag}

The flow times for the region
of interest ($r \le 500$ pc) are short (a few Myr) compared to the lifetime
of the massive stars powering the flow ($\sim 30$ Myr). The properties
of the starburst region will very rapidly approach those of the steady-state
CC solution after any change in the energy and mass injection rate.

Nevertheless, in order to 
apply the CC model 
to M82 we must determine the history (\ie age and star formation
rate as a function of time) of the starburst event in order to derive the
current mechanical energy and mass injection rates and the elemental
abundances in the injected gas. Furthermore, determining 
the degree to which the M82
starburst event can be treated as one the simple limiting cases of
either a single instantaneous burst (SIB) or as continuous star formation
(CSF) is important, in particular for comparison to Starburst99
models \citep{leitherer99}.

Recent observational studies have greatly expanded our knowledge of M82's 
global star formation history since its close encounter with M81 approximately
0.6 --- 1 Gyr ago \citep[see \eg][]{degrijs01_ag}. Nevertheless significant
gaps still remain in our understanding of the history of the nuclear starburst.

Exterior to the central kiloparsec star formation appears to have peaked
$\sim 600$ Myr ago in a burst of comparable intensity to the current
nuclear starburst event, 
but star formation effectively ceased in the disk $\sim 20$ --
30 Myr ago \citep{degrijs01,mayya06}.

Within the central kiloparsec (projected separations from the nucleus of
$r \le 500$ pc) the majority of recent studies have concentrated on the
$\sim$ 200 compact star clusters found there \citep{oconnell95,melo05}.
The brightest $\sim 20$ star clusters typically
appear to have ages $\sim 5$ --- $10$ Myr \citep{satyapal97,mccrady03},
although two massive clusters (M82-F and M82-L) at the edge of this region
have ages of $60\pm{20}$ Myr \citep{l_smith01,l_smith06}. 
\citet{forster03b,forster03a} present a detailed analysis of the 
NIR emission (1 --- 45$\micron$) from a variety of regions within
$r \le 250$ pc, and find that two bursts (of ages $\sim 5$ and $\sim 10$ 
Myr respectively) are required in each region
to simultaneously explain the strength of both nebular 
emission features and CO absorption from Red Super Giant (RSG)
stars. This work, and the estimated ages of
the brightest star clusters, indicates that the nuclear
star formation is not well represented as a single instantaneous starburst 
event. 

Note that the cluster-based studies do not provide a 
complete view of the starburst,
as they are mainly observed in regions of lower-than-average 
extinction\footnote{The typical measured extinction towards the clusters
is $A_{\rm V} \sim 2$ 
--- 6 magnitudes, see \citet{melo05}.} toward the nucleus \citep{l_smith06}.
Nor does this work strongly constrain the star formation rate between
10 --- 60 Myr ago. Old clusters in this age range, of similar mass to
the bright clusters, would be considerably fainter and thus harder to detect
than the $t \sim 10$ Myr-old clusters\footnote{Note 
that this statement, and the previously mentioned cluster ages, 
is based on single star evolutionary models. \citet{mas-hesse99} argue
that if binary systems are taken into account the strongly ionizing phase
of a single coeval cluster can last up to $\sim 30$ Myr after formation,
and that the cluster can simultaneously produce nebular emission and
have a substantial number of RSGs. If this is indeed the case then the
clusters identified as having ages of $\sim 10$ Myr may be as old as
$\sim 30$ Myr.}, due to the rapid 
decrease of the 
ionizing and bolometric output at 
ages $t \ga 10$ Myr. UV spectroscopic 
studies of nearby low-obscuration starbursts 
have shown that the massive star population within the clusters is typically
younger the the field massive star population 
\citep[see \eg][]{tremonti01,chandar05}. The probable cause
for this effect is the gradual dissolution of the massive star clusters
on times scales of $\sim 10$ --- 30 Myr \citep[for a recent review see][and 
references therein]{degrijs07}. We are not aware of any similar UV study
of the field stellar population in M82's nucleus, so there is no direct
evidence for the presence of star formation intermediate in age between
the $t \sim$ 10 Myr-old clusters and the $t \sim 60$ 
Myr-old clusters. Nevertheless
it remains very plausible that significant star formation continued 
in between the two periods.

Another reason to suspect that the nuclear burst has been active for
longer than $\sim 10$ Myr is the dynamical age of the \halpha~cap, a large
nebular feature associated with the superwind at a height $z \sim 12$ kpc
above the mid-plane of the M82 \citep{devine99,lehnert99}. Its true space
velocity is unknown, but assuming a mean velocity of 
$v_{H\alpha} = 600 \kmps$ equivalent
to that of the \halpha~filaments in the lower wind \citep{mckeith95,shopbell98}
it would have taken $\sim 20$ Myr to reach its current location if its
origin was within the disk of the galaxy.

We thus choose to treat M82's nuclear starburst as having experienced
a uniform rate of star formation since the onset of the
burst $t_{\rm burst}$ years ago, with negligible star formation before that
time. We will treat $t_{\rm burst}$ as
a model parameter in the range $10$ --- 50 Myr.

In Table~\ref{tab:burst} we present the bolometric luminosity,
and total SN plus stellar wind energy and mass injection rates (all per
unit star formation rate) for
continuous star formation models calculated using version 5.1 of
Starburst99 population synthesis code \citep{leitherer99}. The star
formation rate is ${\cal{S}} = 1 \Msol \pyr$  for a Salpeter IMF between the
mass limits of 1 --- $100 \Msol$. For a \citet{kroupa01b} IMF between the
mass limits 0.01 --- $100 \Msol$ producing the same total mass of massive
stars (mass range 8 --- $100 \Msol$) as in the Salpeter IMF model, 
the star formation rate is a factor 1.76 times larger. The Geneva stellar
evolutionary tracks for an initial metallicity equal to Solar, and with high
stellar wind mass loss rates, were used.

\begin{deluxetable*}{lrrrrrr}
 \tabletypesize{\scriptsize}%
\tablecolumns{7}
\tablewidth{0pc}
\tablecaption{Continuous star formation models for M82.
        \label{tab:burst}}
\tablehead{
\colhead{Age}
     & \colhead{$L_{\rm bol}/\cal{S}$}
     & \colhead{$\Edot_{\rm SN+SW}/\cal{S}$} 
     & \colhead{$\Mdot_{\rm SN+SW}/\cal{S}$} 
     & \colhead{$\cal{S}$}
     & \colhead{$\Edot_{\rm SN+SW}$} 
     & \colhead{$\Mdot_{\rm SN+SW}$} 
     \\
\colhead{(Myr)}
     & \colhead{($\Lsol \Msol^{-1} \yr$)} 
     & \colhead{($\ergps \Msol^{-1} \yr$)} 
     & \colhead{(\nodata)} 
     & \colhead{($\Msol \pyr$)}
     & \colhead{($\ergps$)} 
     & \colhead{($\Msol \pyr$)} 
     \\
\colhead{(1)}
     & \colhead{(2)} & \colhead{(3)}
     & \colhead{(4)} & \colhead{(5)}
     & \colhead{(6)} & \colhead{(7)}
}
\startdata
10 
   & $9.6 \times 10^{9}$
   & $3.1 \times 10^{41}$ & 0.19
   & 6.1
   & $1.9 \times 10^{42}$ & 1.1 \\
20 
   & $1.1 \times 10^{10}$
   & $5.1\times 10^{41}$ & 0.26
   & 5.1 
   & $2.6 \times 10^{42}$ & 1.3 \\
30 
   & $1.2 \times 10^{10}$
   & $6.5 \times 10^{41}$ & 0.29
   & 4.7 
   & $3.1 \times 10^{42}$ & 1.4 \\
40 
   & $1.3 \times 10^{10}$
   & $7.1 \times 10^{41}$ & 0.31
   & 4.5 
   & $3.2 \times 10^{42}$ & 1.4 \\
50 
   & $1.3 \times 10^{10}$
   & $6.9 \times 10^{41}$ & 0.31
   & 4.4 
   & $3.0 \times 10^{42}$ & 1.3  \\
\enddata
\tablecomments{Column 1: Age since the onset of continuous star formation.
  Column 2:  Starburst radiated bolometric luminosity per unit star
  formation rate. 
  Column 3: Mechanical energy injection rate
  due to SNe and stellar winds per unit star formation rate. 
  Column 4: Mass injection rate due to SNe and stellar winds
  per unit star formation rate.
  Note that the values in Columns 2 -- 4 are specific to a Salpeter
  IMF between the lower and upper mass limits of 1 --- $100 \Msol$.
  Column 5: Star formation rate $\cal{S}$ 
  derived from $L_{\rm IR}/(L_{\rm bol}/\cal{S})$, 
  where $L_{\rm IR}$ is the observed IR luminosity of M82.
  Column 6: The mechanical energy injection rate associated with this SFR.
  Column 7: The mass injection rate associated with this SFR.
  See \S~\ref{sec:theory:applying:sbmag} for details on the 
  stellar evolutionary tracks and stellar wind models used.
}
\end{deluxetable*}

The total 8 --- $1000\micron$ 
IR luminosity of M82 is $L_{\rm IR} = 5.8 \times 10^{10} \Lsol$,
based on the latest re-calibrated 12 -- $100\micron$ 
IRAS fluxes \citep{sanders03}. IR emission dominates M82's spectral
energy distribution, so by equating $L_{\rm IR}$ to the starburst
bolometric luminosity for each assumed age 
yields star formation rates $\cal{S}$ 
between 4 --- $6 \Msol \pyr$ (Salpeter IMF). 
These values and the
 associated SN plus stellar wind energy and mass injection rates 
are tabulated in Table~\ref{tab:burst}. By way of comparison
the formulae in \citet{kennicutt98_review}, which assume CSF
with a large age, give a star formation
rate of ${\cal{S}} = 3.9 \Msol \pyr$ for M82 based on $L_{\rm IR}$,
when corrected to the 1 --- $100 \Msol$ mass range Salpeter IMF
we use.

These star formation rates are not unreasonable. Over the time
scale of interest the net mass of stars formed is at maximum $\sim 50$\%
of the dynamical mass of $\sim 7 \times 10^{8} \Msol$ 
within $r = 250$ pc \citep{gotz90},
even if we assume a Kroupa IMF extending down to a lower mass limit of
$0.01 \Msol$.

\subsubsection{The total size of the starburst region}
\label{sec:theory:applying:sbsize}

The brightest region of diffuse non-thermal 5 GHz radio emission 
in M82 subtends an angular extent of $\sim 43\arcsec \times 6\arcsec$
($\sim 750 \pc \times 105 \pc$),
the same region occupied by the compact radio sources presumed to be
young SN remnants \citep{kronberg85b,muxlow94}. This is similar to
the size of the region occupied by the brightest star clusters
seen in the NIR, $\sim 58\arcsec \times 8\arcsec$ ($\sim 1015 \pc \times
140 \pc$, \citealt{mccrady03}). 
These major-axis angular diameters 
closely match the base of the superwind as measured by the
region of split optical emission lines, which has a diameter of 
$\sim 50\arcsec$ \cite{gotz90}.

If we approximate the starburst region as a flat cylinder of diameter 
$d_{\star}$
and total height $h_{\star}$ then the starburst volume $V_{\star} = 
\pi \, (d_{\star}/2)^{2} \, h_{\star}$ 
and surface area $S_{\star} = \pi d_{\star} (h_{\star} + d_{\star}/2)$ 
in each case is $V_{\star, 1} = 4.6 
\times 10^{7}$ pc$^{3}$, $S_{\star, 1} = 1.1 \times 10^{6}$ pc$^{2}$ 
(5 GHz) and $V_{\star, 2} = 1.1 \times 10^{8}$ pc$^{3}$ , $S_{\star, 2} = 2.1 \times 10^{6}$ pc$^{2}$ (NIR). The equivalent spherical CC model should then
have a starburst region radius of $R_{\star, 1} = 300$ pc or 
$R_{\star, 2} = 405$ pc (Appendix~\ref{app:non_spherical}).

The starburst in M82 is possibly concentrated along the stellar bar 
\citep[see \eg][]{achtermann95,westmoquette07}, and hence 
is more geometrically complex than a uniform disk. Furthermore
star formation may not have occurred simultaneously throughout the full
starburst region, but evidence for inwardly or outwardly propagating
star formation within $r \la 500$ pc of the nucleus remains inconclusive 
\citep{forster03b}. Exploring the full significance of 
these effects requires 3-dimensional hydrodynamical simulations 
that are outside the scope of this paper.

However we can investigate the significance of concentrating the
supernova activity within a smaller disk-like region than the two
considered above. We adopt the starburst disk geometry used in
\cite{ss2000} which has $d_{\star} = 300$ pc and $h_{\star} = 60$ pc. 
This produces
significantly lower starburst volumes and surface areas: 
 $V_{\star, 3} = 4.2 \times 10^{6}$ pc$^{3}$  and 
$S_{\star, 3} = 2.0 \times 10^{5}$ pc$^{2}$. The equivalent spherical
starburst radius is $R_{\star, 3} = 126$ pc.

For the majority of the models we consider we adopt the first starburst region
geometry described above: a disk of diameter $d_{\star} = 750$ pc
and total height $h_{\star} = 105$ pc (or $R_{\star, 1} = 300$ pc for the
scaled CC models). The effect of the size of the starburst region
is explored using the scaled CC model with radii $R_{\star, 2} = 405$ pc
and  $R_{\star, 3} = 126$ pc.



\subsubsection{The effective energy and mass injection rates}
\label{sec:theory:applying:edotmdot}

The most significant contributors to the energy and mass return rates
in a starburst event are core-collapse SNe and the stellar winds from
the massive stars. Stellar winds are thought to contribute an order
of magnitude less mechanical energy than the SNe (assuming equal 
thermalization efficiencies), but they may contribute up to 
$\sim 50$\% of the total mass returned to the ISM 
\citep*{lrd92,leitherer_heckman95}. As described above, we used
the Starburst99 population synthesis code to find the combined
SNe plus stellar wind energy and mass injections rates 
for a continuous burst of star formation
as a function of age. Note that the values
of $\Edot_{\rm SN+SW}$ and $\Mdot_{\rm SN+SW}$ given in  Table~\ref{tab:burst}
have not been corrected for any 
initial mechanical energy losses due to radiation before thermalization
occurs (the thermalization efficiency $\epsilon$) or for mass loading 
by the addition of cold ambient ISM (with a rate $\Mdot_{\rm cold}$).

The thermalization efficiency and degree of mass loading are expected to
be a function of the local gaseous environment the massive stars find
themselves in, as well as the local supernova rate per unit volume and
the density of massive stars. 
Low ambient gas density and/or high SN rate
per unit volume should lead to a higher hot gas filling factor, which
in turns increases the thermalization efficiency by reducing the
initial radiative energy losses \citep{ham90}. However, at very high
stellar densities  
the SN and stellar wind ejecta may itself become dense enough to cause
drastic energy losses, 
\eg within the central regions of super star clusters 
\citep{silich04,tenorio-tagle07}.

To account for these factors we assume that only a fraction $\zeta$
of all the SNe and stellar winds occur in regions where conditions
lead to non-negligible thermalization, and the mean thermalization
efficiency in these regions is what we define as $\epsilon$. 
Essentially $\zeta$ is a modification to the total star formation rate of
the starburst to account for localized spatial variations within the starburst
of the SN and stellar wind feedback. We will refer to $\zeta$ as the 
participation fraction, as it is effectively 
the fraction of the total star formation that is
actually involved in driving the superwind. 
For the remaining fraction 1-$\zeta$ of the
SNe and stellar winds we assume that radiative energy losses are so
significant that no mechanical energy is supplied to the starburst
region, and that the ejecta never becomes part of the outflow. We
treat $\zeta$ as a model variable, and refer the reader to
Appendix~\ref{app:zeta} for a discussion of  
plausible values for $\zeta$ given the gas content of the starburst and
the star cluster mass function.

The net energy injection rate into the hot volume-filling gas 
in the starburst region is
\begin{equation}
\Edot_{\rm tot} = \epsilon \zeta \Edot_{\rm SN+SW}.
\label{equ:edot_tot}
\end{equation} 
The net mass injection rate is 
\begin{equation}
\Mdot_{\rm tot} = \zeta \Mdot_{\rm SN+SW} + \Mdot_{\rm cold} 
      = \beta \zeta \Mdot_{\rm SN+SW},
\label{equ:mdot_tot}
\end{equation}
where the mass loading factor
$\beta$ is a parameter
commonly used to indicate the relative amount of mass loading.
If $\beta = 1$ then the wind is not mass loaded. In this paper we consider
only mass loading that operates within the starburst region, \ie central
mass loading in the terminology adopted by
 \citet{suchkov96} and \citet{ss2000}.

Written as a function of the thermalization efficiency, the mass loading factor
and participation factor, the central temperature, pressure and density are
\begin{equation}
T_{\rm c} = 0.4 \, \frac{\mu m_{\rm H} \, \epsilon \, \Edot_{\rm SN+SW}}{
  k \, \beta \, \Mdot_{\rm SN+SW}},
\label{equ:final_tc}
\end{equation}
\begin{equation}
P_{\rm c} = 0.118 \, \frac{ \epsilon^{1/2} \, \beta^{1/2} \, \zeta \,  
  \Edot_{\rm SN+SW}^{1/2} \, \Mdot_{\rm SN+SW}^{1/2}}{ R_{\star}^{2}},
\label{equ:final_pc}
\end{equation}
and
\begin{equation}
\rho_{\rm c} = 0.296 \, \frac{ \zeta \, \beta^{3/2} \,  
  \Mdot_{\rm SN+SW}^{3/2}}{\epsilon^{1/2} \,  
  \Edot_{\rm SN+SW}^{1/2} 
  \, R_{\star}^{2}}.
\label{equ:final_rhoc}
\end{equation}

The central temperature is thus linearly proportional to the ratio of
the mechanical energy thermalization efficiency to the mass loading factor,
and independent of the size of the starburst region $R_{\star}$
or participation factor $\zeta$. 
Thus it is an excellent diagnostic tool for quantifying the effects of 
feedback, provided that this temperature can be accurately measured.

The terminal velocity of the wind fluid
\begin{equation}
v_{\infty} = \left( \frac{2 \, \epsilon  \, \Edot_{\rm SN+SW}}{\beta \Mdot_{\rm SN+SW}} \right)^{1/2} \equiv \left( \frac{5 k T_{\rm c}}{\mu m_{\rm H}} \right)^{1/2}
\label{equ:vinf}
\end{equation}
is also a function of $\epsilon/\beta$. An observational determination
of $v_{\infty}$ is beyond the capabilities of current
X-ray observatories, but the  X-ray Microcalorimeter 
Spectrometer (XMS) instrument on the future International X-ray 
Observatory (IXO) will provide the high resolution
non-dispersive X-ray spectroscopy necessary to directly measure the
velocity of the hot phases in superwinds.

\subsubsection{The metal abundance within the wind fluid}
\label{sec:theory:applying:z}

The mixed SN ejecta and stellar wind material will be enriched in
heavy elements compared to the cooler ambient ISM within the starburst
region. As the emissivity of the thermal X-ray 
emission from a hot plasma depends
strongly on the metal abundance of the hot plasma it is necessary to consider
both SN-related enrichment and dilution via mass loading.

If we assume mass-loading is associated with efficient mixing, then
the elemental abundance  $Z_{i, {\rm WF}}$ of an element $i$ in the wind fluid is
\begin{equation}
Z_{i, {\rm WF}} = \frac{Z_{i, {\rm SN+SW}} \zeta \Mdot_{\rm SN+SW} 
        + Z_{i, {\rm cold}} \Mdot_{\rm cold}}{\zeta \Mdot_{i, {\rm SN+SW}} 
        + \Mdot_{\rm cold}} 
      = \frac{Z_{i, {\rm SN}+SW} + (\beta - 1) Z_{i, {\rm cold}}}{\beta}
\label{equ:zwf}
\end{equation}
where $Z_{i, {\rm SN+SW}}$ is the abundance of the element in the 
SN plus stellar wind ejecta and $Z_{i, {\rm cold}}$ is the abundance
in the ambient ISM, and we have made use of Equation~\ref{equ:mdot_tot} which
defines $\beta$.

The primary elemental coolants
 for plasmas with temperature $T \ga 10^{5}$ K are
oxygen, neon and iron \citep[\eg see][]{sutherland93}. Iron is of particular
importance for this work given the $E\sim 6.7$ keV and $E\sim 6.9$ keV
lines from helium-like and hydrogen-like iron ions. Elements such as S, 
Ar and Ca also produce weaker X-ray lines at energies $E > 2$ keV that may also
serve as a diagnostic of the wind-fluid.

\citet{limongi07}
present the IMF-averaged SN ejecta (both SN II and SN Ib/c) 
element abundances with respect to 
Solar\footnote{All abundances quoted in this paper are relative to the
scale defined in \citet{anders89}.}. Their stellar evolution models include
mass loss due to stellar winds, but the quoted yields do not include the
hydrogen and lighter elements returned to the ISM by the winds. They
find $Z_{\rm Fe, SN} \sim 7 Z_{Fe, \odot}$, $Z_{\rm O, SN} \sim 
6 Z_{O, \odot}$, $Z_{\rm Ne, SN} \sim 10 Z_{Ne, \odot}$, 
$Z_{\rm S, SN} \sim 5 Z_{S, \odot}$, 
$Z_{\rm Ar, SN} \sim 4 Z_{Ar, \odot}$ and 
$Z_{\rm Ca, SN} \sim 3 Z_{Ca, \odot}$. Note the low O/Fe abundance
ratio compared to the IMF-averaged yields presented in 
\citet{gibson97b}, which is most probably a
consequence of wind mass loss in the stellar evolution models used by
\citet{limongi07}. 

The time-averaged nature of 
an IMF-averaged yield is appropriate for continuous star formation with an
age of $t \ga 20$ Myr, but the abundances in the ejecta for a younger 
or instantaneous burst may differ from that in \citet{limongi07}.
To investigate this, and the effect of using a different set of SN yields, 
we computed the abundances of the combined SN ejecta and stellar wind 
material from the yields given by Starburst99. Starburst99
includes wind yields by combining the stellar evolutionary wind mass loss
rates with observed stellar surface abundances. SN yields are based
on \citet{woosley95} for SN type II, but note that the the stellar evolution
models that \citeauthor{woosley95} 
used did not in general include wind mass loss. 
Starburst99 also assumes that stars that become Wolf-Rayet 
stars (initial mass $M_{\star} \ga 33 \Msol$) are assumed to 
explode as SN type Ib/c and contribute no metals to the final yields. 
For a continuous star formation model the SN plus stellar wind
ejecta abundances predicted by Starburst99
are relative constant for ages $t \ge 15$ Myr after the
onset of the burst, at levels of 
$Z_{\rm Fe, SN+SW} \sim 4 Z_{Fe, \odot}$, $Z_{\rm O, SN+SW} \sim 
6 Z_{O, \odot}$, and $Z_{\rm S, SN+SW} \sim 14 Z_{S, \odot}$
(Starburst99 does not calculate the yields of Ne, Ar and Ca).

Given the uncertainties we will adopt 
$Z_{\rm Fe, SN+SW} = 5 Z_{\rm Fe, \odot}$ for these models. For computational
convenience we will further assume the same abundance relative to
Solar for the other elements.

The stellar and \ion{H}{2} region O and Ne abundances in M82
are Solar or higher \citep{achtermann95,origlia04,smith06}.
The are very few measurements of iron abundances, the only one we
are aware of is that of \citet{origlia04} who
derive a stellar iron abundance of $Z_{\rm Fe} =
0.46^{+0.26}_{-0.17} Z_{\rm Fe, \odot}$ from NIR spectroscopy.
The total baryonic mass of M82 is $\sim 2 \times 10^{10} \Msol$
based on its peak rotational velocity \citep[see][]{strickland04a}.
Galaxies of this mass have a mean nebular oxygen abundance of
$\sim 1.2 Z_{\rm O, \odot}$ (on the \citealt{anders89} scale),
based on the galaxy mass-metallicity relationship \citep{tremonti04}. 
Thus the measured metallicity of the stellar population and the 
warm ISM in M82 are not unusual given its mass, nor do they suggest
that the ambient ISM has been significantly enriched with nucleosynthetic
products from the current starburst.
We adopt $Z_{\rm Fe, cold} = 1 Z_{\rm Fe, \odot}$ for our models, again assuming
the same value for all other elements.

\begin{figure*}[!ht]
\plotone{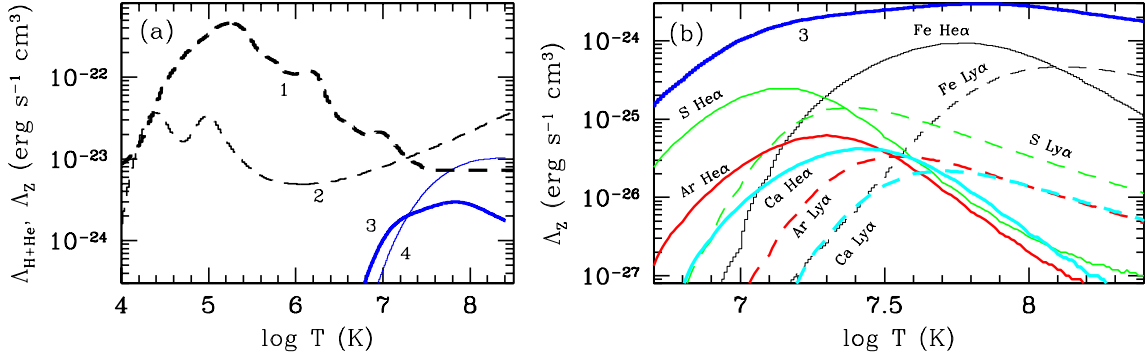}
  \caption{(a) The broad band cooling functions used in the model 
  calculations. The total
  cooling due to metals for unit metal abundance with respect to Solar
  abundance (1)
  and cooling from hydrogen and helium only (2) for non-equilibrium 
  cooling are both from \citet{sutherland93}. The APEC plasma code
  \citep{apec01} was used to calculate the emission in the hard 
  X-ray energy band ($E=2$ -- 8 keV) from metals (3) and hydrogen 
  plus helium (4), assuming collisional ionization equilibrium.
  (b) X-ray cooling curves per unit metal abundance
  for helium-like and hydrogen-like ions of S, Ar, Ca and Fe in comparison to
  the 2--8 keV cooling curve for all metals (3). 
  }
  \label{fig:emis}
\end{figure*}

\subsection{Broad-band cooling and X-ray line emission}
\label{sec:theory:applying:emission}

The luminosity per unit volume $\cal{L}$ ($\ergps \pcc$) due to the
thermal emission from an element of gas of density $\rho$, 
temperature $T$ and metal abundance $Z$ is
\begin{equation}
{\cal L} = n_{e} n_{\rm H} \Lambda(T, Z) 
        = \frac{\chi_{\rm H} (1+ \chi_{\rm H})}{2 m_{\rm H}^{2}}
          \, \rho^{2} \,  \Lambda(T, Z),
\label{equ:l_per_unit_vol}
\end{equation}
where $\chi_{\rm H} \approx 0.71$ is the mass fraction of hydrogen for
a Solar abundance plasma, $m_{\rm H}$ is the mass of a hydrogen atom
and $\Lambda(T,Z)$ is the cooling function for the specific energy band
or line transition being considered.

For broad-band X-ray emission, or the total cooling rate, the cooling function
for a hot optically thin plasma may be split into a metallicity-independent
component $\Lambda_{\rm H+He}(T)$  
and metallicity-dependent component $\Lambda_{Z}(T)$ (evaluated for
Solar abundance), such that
$\Lambda(T, Z) = \Lambda_{\rm H+He}(T) + Z \times \Lambda_{Z}(T)$,
where $Z$ is the metal abundance with respect to the Solar. This
assumes that the relative metal abundances are Solar, as otherwise
the contribution from each element must be considered separately.

To assess the total cooling rate we use the zero-field
non-equilibrium cooling functions from \citet{sutherland93}. We calculated
the metallicity-dependent cooling $\Lambda_{Z}$ from their published Solar
and zero metallicity cooling functions. At 
temperatures above $T \sim 5 \times 10^{6}$ K (typical of the vast majority of
the models we will consider) the difference between
equilibrium and non-equilibrium cooling functions is negligible.

To calculate the cooling functions for 
the $E=2$ --- 8 keV hard X-ray band and the He$\alpha$ and
Ly$\alpha$ like lines of S, Ar, Ca and Fe we used
version 1.3.1 of APEC hot plasma code (which assumes collision 
ionization equilibrium), accessed using the {\sc Xspec}
spectral fitting program \citep{xspec_ref,apec01}. 
For the line
emission from any specific element $i$ the cooling function is an 
exactly linear function of the metallicity $Z_{i}$, \ie 
$\Lambda(T, Z_{i}) = Z_{i} \times \Lambda_{Z}(T)$. 


In Fig.~\ref{fig:emis} we plot the total and 2 -- 8 keV cooling functions
(split into their metallicity-independent and metallicity-dependent 
components) that we use in our calculations. 
The line-specific cooling functions for hydrogen-like and helium-like 
sulphur, argon, calcium and iron for unit metal abundance are also shown. 

The X-ray line emissivities we use apply to hot gas in collisional
ionization equilibrium (CIE). In our the calculations using the scaled
CC model  we evaluate the following expression in each radial shell as 
we numerically integrate outwards:
\begin{equation}
\eta = \frac{n_{e}}{v | \frac{d \log T}{dr}|} 
\label{equ:eta}
\end{equation}
If the ionization parameter $\eta \ga 10^{12} {\rm cm}^{-3} {\rm s}$
then the plasma is likely to be in CIE \citep{masai02}. If
$\eta < 10^{12} \pcc$ s then the outflowing gas is over-ionized 
and in recombining
non-equilibrium. \citeauthor{masai02} show that the broad band X-ray emissivity
of such a recombining plasma is less than that of a CIE plasma. 
The ionization parameter $\eta$ scales with
thermalization efficiency and mass loading proportionally as
$\beta^{2}/\epsilon$, as increasing the mass loading or decreasing the
thermalization efficiency acts both to increase the plasma density and
decrease the outflow velocity.

For parameters appropriate to M82 we find that the plasma is typically
in CIE at radii $r \la R_{\star}$ even when $\epsilon = \beta = 1$. 
Even if we integrate the emission
out to radii $\gg R_{\star}$ the total luminosity is dominated by emission
from within $R_{\star}$, justifying our use of CIE emissivities. The radial
variation in $\eta$ and the cumulative emission integral $EI(<R)$
shown in Fig~\ref{fig:radial_solution}c demonstrate this for one set of
possible model parameters. This result is consistent with the 
alternative ionization equilibrium and Coulomb relaxation timescale 
calculations presented in Paper I, which also indicated that the hot
plasma within the starburst region should be in ionization equilibrium.

For a single temperature plasma the luminosity is the product of the 
emission integral $EI$ with the cooling function $\Lambda$.
The cumulative volume emission integral is defined as 
$EI(<R) = \int_{0}^{R} 4\pi n_{e} n_{\rm H} R^{2} dR$. 
For a CC-style radial wind this
is dominated by material within the relatively dense starburst region,
as shown in Fig~\ref{fig:radial_solution}c. Thus the total emission
integral is proportional to
\begin{equation}
EI \propto \rho^{2} \, V_{\star} 
   \propto \frac{\zeta^{2} \, \beta^{3}}{\epsilon \, R_{\star}} \,
   \frac{\Mdot_{\rm SN+SW}^{3}}{\Edot_{\rm SN+SW}},
\label{equ:ei}
\end{equation}
which again increases as either the mass loading factor $\beta$ is
increased, or as the thermalization efficiency $\epsilon$ is decreased.
However such variations in $\beta$ and/or $\epsilon$ 
are also associated with a decrease in the central temperature,
and for increased mass loading a decrease in the mean metal abundance,
so that the net luminosity can ultimately decrease even as
increasing $\beta$ or decreasing $\epsilon$ monotonically increase the
net emission integral.

\subsection{Observational constraints}
\label{sec:theory:applying:constraints}

The primary aim of this project is to assess
the range of starburst properties that 
are consistent with the hard X-ray Fe-line-emitting plasma in M82, and
which models are deemed acceptable depends on which constraints are
used --- this is unavoidable. These are not an arbitrary set of observational
and conceptual constraints but have been chosen based on realistic
physical expectations.

Based on the results and discussion
presented in \S~\ref{sec:observations} we adopt the
following as the full observational
constraints on the properties of the wind fluid within a radius 
of 500 pc of the center of M82. 

\begin{itemize}
\item The total X-ray luminosity of the wind fluid
  in the $E=2$ -- 8 keV energy band must be
  less than or equal to the broad band luminosity estimated from 
  the ACIS-S observation for the diffuse hard X-ray emission
  (Table~\ref{tab:paper1}), \ie 
  $\log L_{\rm 2-8 keV} \le 39.65$. 
\item The Fe He$\alpha$ line luminosity must be consistent with 
  the observed value given in Table~\ref{tab:paper1}.
  In order to be conservative in our estimation of
  the constraints the observations place on the thermalization efficiency
  and mass loading, \ie to accept the largest number of models matching
  the observational constraints, we adopt the \emph{weakest} constraints 
  on the allowed value of $L_{\rm Fe He\alpha}$, \ie using the ACIS-S 
  observation rather than the better constrained values from the
  {\it ACIS-I} or {\it XMM-Newton} data sets. Models producing Fe line
  luminosities within 68.3\% confidence
  region from the ACIS-S observation, $37.96 \le 
  \log L_{\rm Fe He\alpha} \le 38.39$, will be judged as matching the
  constraint on the $E\sim6.7$ keV line emission.
  This measurement of $L_{\rm Fe He\alpha}$ is slightly larger than,
  but still consistent with the
  line luminosity estimated from the ACIS-I and XMM-Newton observations.
\item The Fe Ly$\alpha$ luminosity must be less than or equal to the 
  observationally-determined upper limits, of which the best determined
  value is from the {\it  XMM-Newton} observation. We specify that
  $L_{\rm Fe Ly\alpha}/L_{\rm Fe He\alpha}$ should be $\le 0.34$ 
  for a model to be judged to be consistent with 
  the observational constraints (Table~\ref{tab:paper1}). 
\item The luminosity of the predicted He$\alpha$ and Ly$\alpha$-like lines
  of S, Ar and Ca must be less than or equal to the observed values and/or
  99\% upper limits. Here the best observational constraints are from
  the {\it XMM-Newton} data (Table~\ref{tab:s_ar_ca}).
  Acceptable models must then satisfy all of the
  following: 
  $\log L_{\rm S He\alpha} \le 38.19$, 
  $\log L_{\rm S Ly\alpha} \le 37.43$,
  $\log L_{\rm Ar He\alpha} \le 37.53$, 
  $\log L_{\rm Ar Ly\alpha} \le 37.26$,
  $\log L_{\rm Ca He\alpha} \le 37.27$, and 
  $\log L_{\rm Ca Ly\alpha} \le 37.14$.
  Upper limits are used as the S, Ar and Ca-line-emitting plasma is not
  necessarily the same phase as the Fe-line-emitting plasma.
\end{itemize}

Models that (a) have total radiative losses less than 30\% of the net energy
injection rates (see Appendix~\ref{app:radiative_case}) 
and (b) predict X-ray luminosities consistent with the 
full set of constraints given above are considered as having successfully
matching the full set of constraints.

We will consider the effect of using a different set of 
observational constraints. These alternatives deliberately ignore
what we know about the iron line emission (the iron lines are not
used as constraints) and instead attempt to fully
account for all of either the broad-band hard X-ray luminosity or
the detected S, Ca and Ar line luminosities:
\begin{itemize}
\item The total X-ray luminosity of the wind fluid
  in the $E=2$ -- 8 keV energy band must be within the 3$\sigma$
  envelope given in Table~\ref{tab:paper1} for the model to considered
  successful, \ie
  $39.59 \le \log L_{\rm 2-8 keV} \le 39.71$. In addition the
  the total radiative losses must be less than 30\% of the net energy
  injection rate.
  We shall refer to these constraints as the broad-band-only constraints.
\item The predicted line luminosities for the S He$\alpha$, S Ly$\alpha$,
  Ar He$\alpha$ and Ca He$\alpha$ must be within $\pm{3}\sigma$ of the
  values given in Table~\ref{tab:s_ar_ca} for the best fit to  
  the {\it XMM-Newton} spectrum. 
  The allowed limits on the Ar Ly$\alpha$ and Ca Ly$\alpha$
  line luminosities are the same as given for the full set of constraints.
  In addition the model must predict broad band and total radiative 
  energy losses less than the upper limits also used in the full set of
  constraints:
  $\log L_{\rm 2-8 keV} \le 39.65$ and total radiative losses 
  less than 30\% of the net energy injection rate.
  We shall refer to these constraints as the SArCa constraints.
\end{itemize}

We allow larger ranges in the broad-band or line luminosity ($\pm{3}\sigma$)
as we found that in practice that it is relatively hard to find regions
of parameter space where models can satisfy the broad-band-only or SArCa 
constraints.

\section{Simulations and the results thereof}
\label{sec:results}

\begin{figure*}[!t]
\epsscale{0.9}
\plotone{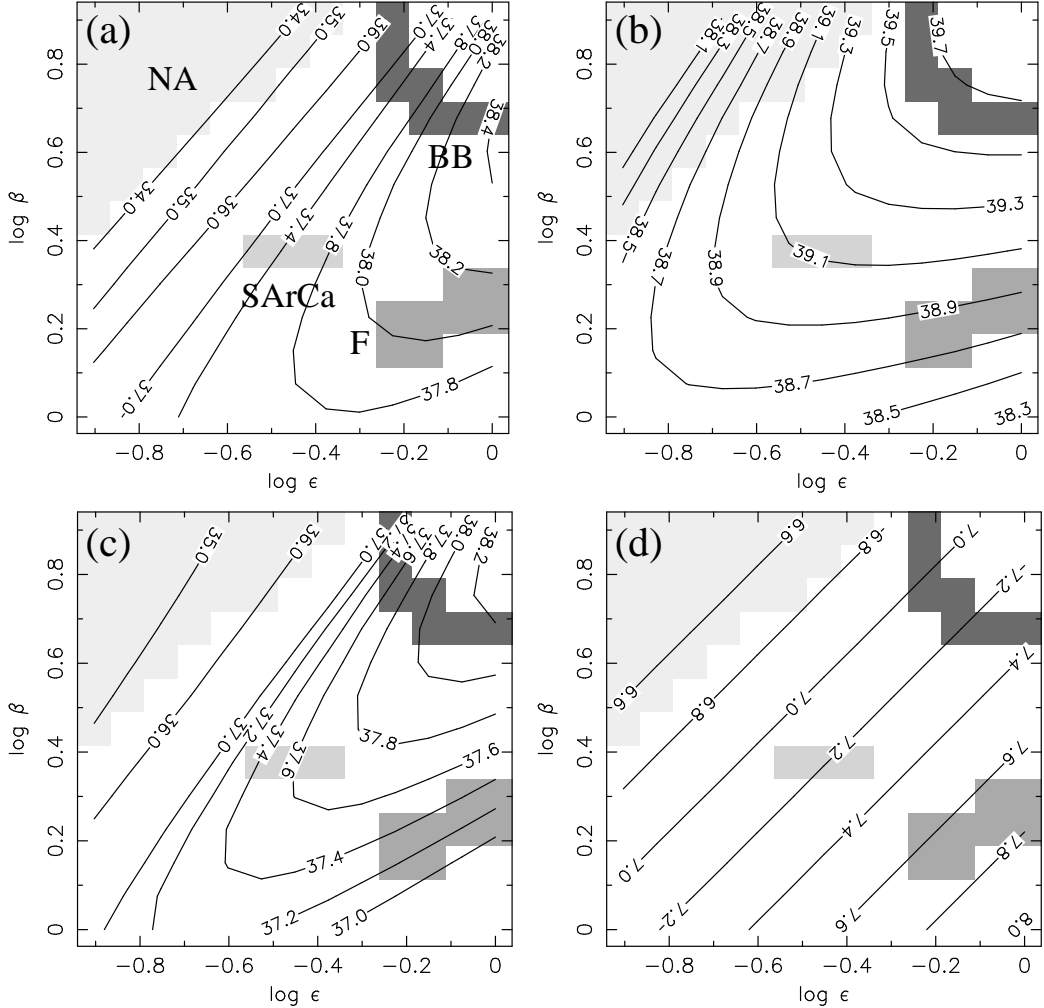}
  \caption{Predicted luminosities and temperatures as a function of
  the input thermalization efficiency $\epsilon$
  and mass loading factor $\beta$, as predicted using
  the 2-dimensional hydrodynamical simulations with parameter set A 
  (see \S~\ref{sec:model:epsilon_and_beta}).
   (a) The logarithm of the Fe He$\alpha$ luminosity.
  The grey-shaded region labeled F shows the region of parameter space
  that matches the full set of observational constraints discussed in
  \S~\ref{sec:theory:applying:constraints}. The region labeled BB denotes
  the region satisfying the broadband-only constraints, while the region
  labeled SArCa satisfies the S, Ar, and Ca line constraints. The region
  labeled NA denotes the region of parameter space where the adiabatic
  assumption ceases to the valid.
  (b) As (a), except the labeled 
  contours show the logarithm of the luminosity 
  in the $E=2$ -- 8 keV energy band.
  (c) As (a), except the labeled contours show
   the logarithm of the S Ly$\alpha$
  line luminosity.
  (d) As (a), except the contours show the logarithm of the central 
  plasma temperature $T_{\rm c}$ in Kelvin.}
  \label{fig:new_medium_vh1}
\end{figure*}

\subsection{Acceptable values of $\epsilon$ and $\beta$.}
\label{sec:model:epsilon_and_beta}

Mass loading factors of $\beta \sim 10$ or thermalization efficiencies
of $\epsilon \sim 0.1$ mark the practical upper and lower limits on these
parameters for a M82-like starburst. Each of these values on its own will
reduce the central temperature to $T_{\rm c} \sim 10^{7}$ K, a temperature
at or below which the emissivity of the Fe He$\alpha$ emission drops
dramatically (Fig.~\ref{fig:emis}).

To explore this region of the mass loading and thermalization
efficiency parameter space
we adopt the following values for the other important model parameters:
$\zeta =1$, a disk-like starburst with $d_{\star} = 750$ pc and 
$h_{\star} = 105$ pc ($R_{\star} = 300$ pc), $Z_{\rm SN+SW} = 5 \Zsol$ and
$Z_{\rm cold} = 1 \Zsol$. We treat
the starburst as continuous star formation at a rate of
${\cal{S}} = 4.7 \Msol \pyr$ beginning 30 Myr ago,
which leads to $\Edot_{\rm SN+SW} = 3.1 \times 10^{42} \ergps$ and
 $\Mdot_{\rm SN+SW} = 1.4 \Msol \pyr$. We shall refer to this combination
of input model parameters as parameter set A 
(see Table~\ref{tab:parameter_sets}).

\begin{deluxetable*}{lcrcrcr}
 \tabletypesize{\scriptsize}%
\tablecolumns{7}
\tablewidth{0pc}
\tablecaption{Combinations of starburst model parameters explored in the
  simulations.
        \label{tab:parameter_sets}}
\tablehead{
\colhead{Parameter Set}
     & \colhead{SF Mode/Age}
     & \colhead{$\Edot_{\rm SN+SW}$} 
     & \colhead{$\Mdot_{\rm SN+SW}$} 
     & \colhead{$R_{\star}$}
     & \colhead{$Z_{\rm SN+SW}$} 
     & \colhead{$\zeta$} 
     \\
\colhead{\nodata}
     & \colhead{\nodata} 
     & \colhead{($\ergps$)} 
     & \colhead{($\Msol \pyr$)} 
     & \colhead{(pc)}
     & \colhead{($Z_{\odot}$)} 
     & \colhead{\nodata}
     \\
\colhead{(1)}
     & \colhead{(2)} & \colhead{(3)}
     & \colhead{(4)} & \colhead{(5)}
     & \colhead{(6)} & \colhead{(7)}
}
\startdata
A & CSF / 30 Myr & $3.1 \times 10^{42}$ & 1.4 & 300.0 & 5.0 & 1.00 \\
B & CSF / 10 Myr & $1.9 \times 10^{42}$ & 1.1 & 300.0 & 5.0 & 1.00 \\
C & SIB / 10 Myr & $4.5 \times 10^{42}$ & 2.2 & 300.0 & 5.0 & 1.00 \\
D & CSF / 30 Myr & $3.1 \times 10^{42}$ & 1.4 & 125.5 & 5.0 & 1.00 \\
E & CSF / 30 Myr & $3.1 \times 10^{42}$ & 1.4 & 404.9 & 5.0 & 1.00 \\
F & CSF / 30 Myr & $3.1 \times 10^{42}$ & 1.4 & 300.0 & 2.5 & 1.00 \\
G & CSF / 30 Myr & $3.1 \times 10^{42}$ & 1.4 & 300.0 & 7.5 & 1.00 \\
H & CSF / 30 Myr & $3.1 \times 10^{42}$ & 1.4 & 300.0 & 5.0 & 0.71 \\
I & CSF / 30 Myr & $3.1 \times 10^{42}$ & 1.4 & 300.0 & 5.0 & 0.50 \\
\enddata
\tablecomments{Column 1: Adopted name of combinations of model parameters.
  Column 2: The mode of star formation assumed, either continuous star
  formation (CSF) or an single instantaneous burst (SIB) beginning or occurring
  the specified time before the present.
  Columns 3 and 4: The mechanical energy and mass injection rates 
  (See \S~\ref{sec:theory:applying:sbmag}) associated with the assumed
  star burst.
  Column 5: The effective starburst radius adopted in calculations using 
  scaled CC model.
  Column 6: The assumed metal abundance of heavy elements relative to Solar
  (see \S~\ref{sec:theory:applying:z}).
  Column 7: The assumed participation fraction 
  (see \S~\ref{sec:theory:applying:edotmdot} or Appendix~\ref{app:zeta}).
}
\end{deluxetable*}

We initially computed models covering the full range of $\epsilon:\beta$
parameter space. To prevent the total number of simulations from becoming 
excessive we considered only a limited number of distinct values 
for $\epsilon$ and $\beta$,
specifically $\epsilon = 2^{-N}$ and $\beta = 2^{N}$, 
where N ranges from 0 to 3 in increments of 0.25, resulting in 
the computation of 169 models using the 2-dimensional hydrodynamical code
described in \S~\ref{sec:hydromodel}. 

We find that only limited regions of parameter space yield models that
satisfy any one of the sets of observational constraints we chose to apply 
(Fig.~\ref{fig:new_medium_vh1}). Furthermore, the regions in 
$\epsilon$:$\beta$ parameter space satisfying the different sets of constraints
are well separated from one another.

Models that satisfy the full set of observational constraints (labeled F
in Fig.~\ref{fig:new_medium_vh1}) have moderately high thermalization
efficiencies ($\log \epsilon \ga -0.3$) and moderately low but non-negligible
mass loading ($0.15 \la \log \beta \la 0.3$). 
These models reproduce the observed
Fe line fluxes and limits without overproducing S, Ar or Ca line emission, but
in doing so they can only account for a small fraction ($\sim 10$ -- $25$\%)
of the total observed
diffuse luminosity of $\log L_{\rm 2-8 keV} \sim 39.6$  in 
the $E=2$ -- 8 keV energy band (Fig.~\ref{fig:new_medium_vh1}b).

The range of central thermal pressures found in these matching models, 
$P_{\rm c}/k = (1.4$ -- $2.5) \times 10^{7}$ K cm$^{-3}$
(given in Table~\ref{tab:allowed_regions}) is consistent with
estimates of the pressure within the starburst region obtained
from optical spectroscopy: $P/k \sim (0.5$ -- $3) \times 10^{7}$  K cm$^{-3}$
\citep{ham90,smith06,westmoquette07}. 

To match the observed S, Ar and Ca line fluxes using a wind model 
(the region labeled SArCa), models require a lower 
thermalization efficiency and a comparable or slightly higher mass loading
factor ($-0.7 \la \epsilon \la -0.4$, $\log \beta \sim 0.4$). 

To reproduce the observed diffuse X-ray luminosity in the $E=2$ -- 8 keV 
energy band with thermal emission alone  requires both high 
thermalization efficiency and high mass-loading factors: 
$\log \epsilon \ga -0.3$ and $\beta \ga 5$ (the region labeled 
BB in Fig.~\ref{fig:new_medium_vh1}). Models in this region of
parameter space fail to match the full set of observational constraints
because they typically predict
more S, Ar or Ca emission (the S Ly$\alpha$ luminosity is shown in
Fig.~\ref{fig:new_medium_vh1}c) than is observed when the Fe line luminosity
is correctly reproduced, or too little Fe He$\alpha$ emission when they
predict less than or equal to the amount of S, Ar or Ca line emission observed.

Thus we can only account for $\la 50$\% of the observed $E=2$ -- 8 keV
diffuse X-ray luminosity with $R=500$ pc with the sum of the predicted 
thermal emission from the starburst region and the Inverse Compton
X-ray emission (Paper I). We conclude that either our estimate of the 
Inverse Compton luminosity is in error, or there must be an additional 
non-thermal or quasi-thermal source of hard X-ray continuum emission within 
the starburst region \citep[\eg][]{dogiel02,masai02}. Solving this enigma
will require higher resolution spectroscopy
to establish the shape of the continuum and
the ionization state of the line emission.

Models where the total radiative losses are higher than 30\% of the
energy injection rate, and the adiabatic assumption is no longer valid, 
are labeled NA in  Fig.~\ref{fig:new_medium_vh1}. Note that for these 
models the temperatures and X-ray luminosities we predict are not valid.
The
location of this region in $\epsilon:\beta$ space in consistent with 
the discussion in Appendix~\ref{app:radiative_case}. 
For M82's star formation rate and starburst region size we 
find that as long as $\zeta \beta^{3}/\epsilon^{2} \la 1000$ then
the adiabatic assumption is applicable to either a disk-like or spherical
starburst.

Parameters that give rise to highly radiative winds can be rejected even 
though we can not predict their exact X-ray luminosities. Cooling is so
efficient under such conditions that the central temperature $T_{\rm c} \ll 
10^{7}$ K and there is no possibility of Fe He$\alpha$ emission
\citep[\eg][]{tenorio-tagle07}.

\begin{figure*}[!t]
\epsscale{0.9}
\plotone{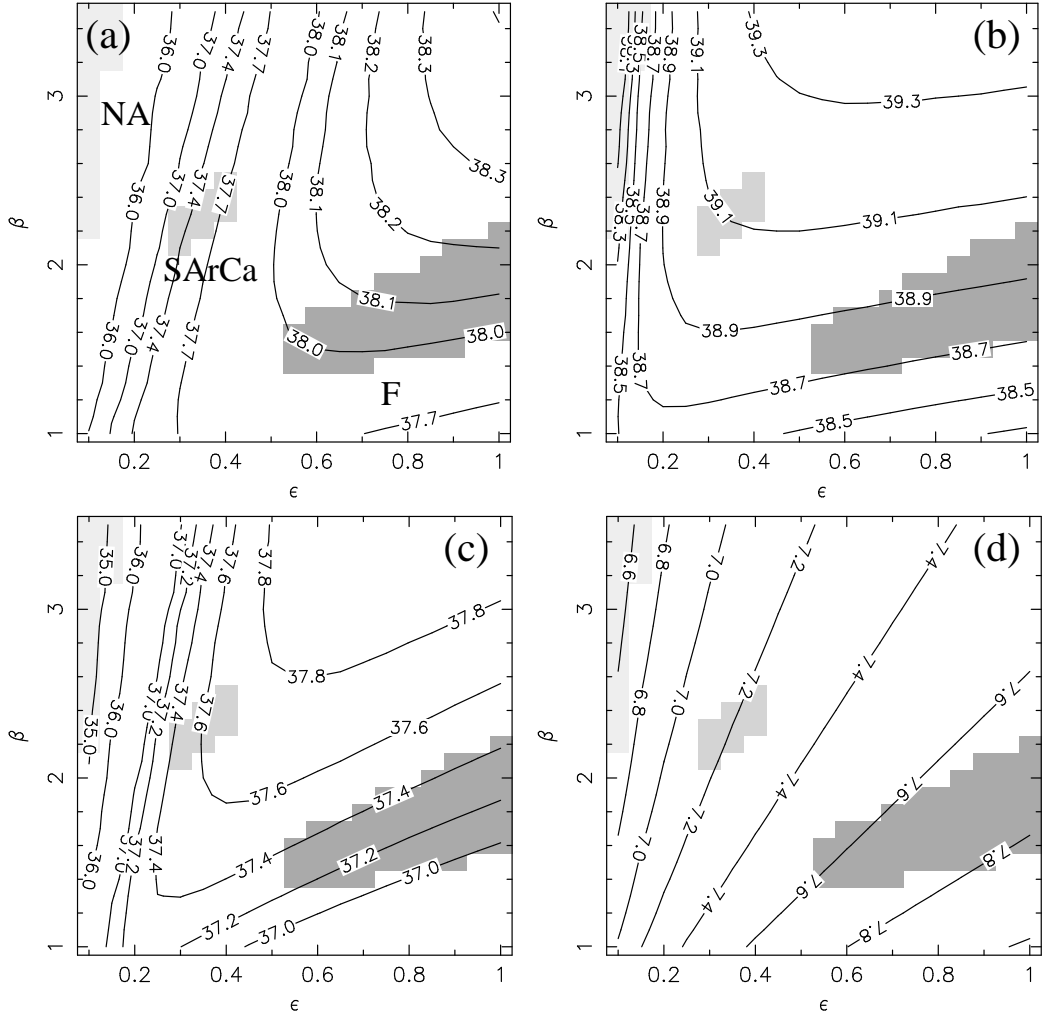}
  \caption{As Fig.~\ref{fig:new_medium_vh1}, except using a finer
  grid of models 
  distributed linearly in a smaller region of $\epsilon:\beta$ parameter 
  space. The numbered contour levels are of log $L_{\rm Fe He\alpha}$ (a),
  log $L_{\rm 2-8 keV}$ (b), log $L_{\rm S Ly\alpha}$ (c) 
  and log $T_{c}$ (d).}
  \label{fig:new_fine_vh1}
\end{figure*}

To further refine our knowledge of the exact region of the allowed
$\epsilon:\beta$ parameter space for M82's starburst we
ran a further set of full hydrodynamical simulations 
a subset of the full  $\epsilon:\beta$ grid described above.
We used a linearly-spaced grid, varying $\epsilon$ between 0.1 and 1.0 in
steps of $\Delta \epsilon = 0.05$ and $\beta$ between 1.0  and 3.0 in
steps of $\Delta \beta = 0.1$ (a total of of 494 models). 

The predicted regions of allowed parameter space for parameter set A
are shown in Fig.~\ref{fig:new_fine_vh1}. This figure also demonstrates that
the exact shape of the allowed region using the full set of constraints
is most-strongly influenced by the combination of the constraints on the
Fe He$\alpha$ line and the upper limit on the S Ly$\alpha$ 
line\footnote{For this reason we have
not provided plots of the S He$\alpha$, Ar or Ca line luminosities as a 
function of $\epsilon$ and $\beta$.}. For a given value of the 
thermalization efficiency $\epsilon$
the lower boundary of acceptable mass loading factors $\beta$
is set by requirement to achieve sufficient Fe He$\alpha$ emission
(the necessary condition is $37.96 \le \log L_{\rm Fe He\alpha} \le 38.39$),
while the maximum $\beta$ is set by the requirement not to exceed the observed
S Ly$\alpha$ luminosity ($\log L_{\rm S Ly\alpha} \le 37.43$). 

For a given value of $\epsilon$ the highest allowed $\beta$ is a less
likely solution than a lower allowed value of $\beta$. The highest allowed
values of $\beta$ produce the entire observed S Ly$\alpha$ flux in the wind
fluid, without leaving leeway for S Ly$\alpha$ emission from the
soft X-ray-emitting plasma within the starburst region.

Were we to change the S Ly$\alpha$ constraint from an upper limit to a specific
range in values around the observed line luminosity the resulting allowed 
region would
be a narrow strip along what is currently the upper $\beta$ margin of the full
allowed region. Adopting such a constraint would be somewhat ad-hoc, 
given that the S He$\alpha$ and Ar He$\alpha$ lines are
equally well detected as the S Ly$\alpha$ line (Table~\ref{tab:s_ar_ca}).
Furthermore, an attempt to match the S He$\alpha$, Ly $\alpha$ and 
Ar He$\alpha$ line
fluxes, rather than treating them as upper limits, simultaneously with the Fe 
He$\alpha$ line flux leads to no acceptable models being found. Ultimately this
is because the S, Ar and Ca line fluxes and upper limits are indicative of
a plasma temperature $\log T \la 7.2$ (see Table~\ref{tab:temp_estimates} or 
the SArCa region on Fig.~\ref{fig:new_fine_vh1}d), yet such a low temperature
plasma produces too little Fe He$\alpha$ emission to match the observations.

\begin{figure*}[!t]
\epsscale{0.9}
\plotone{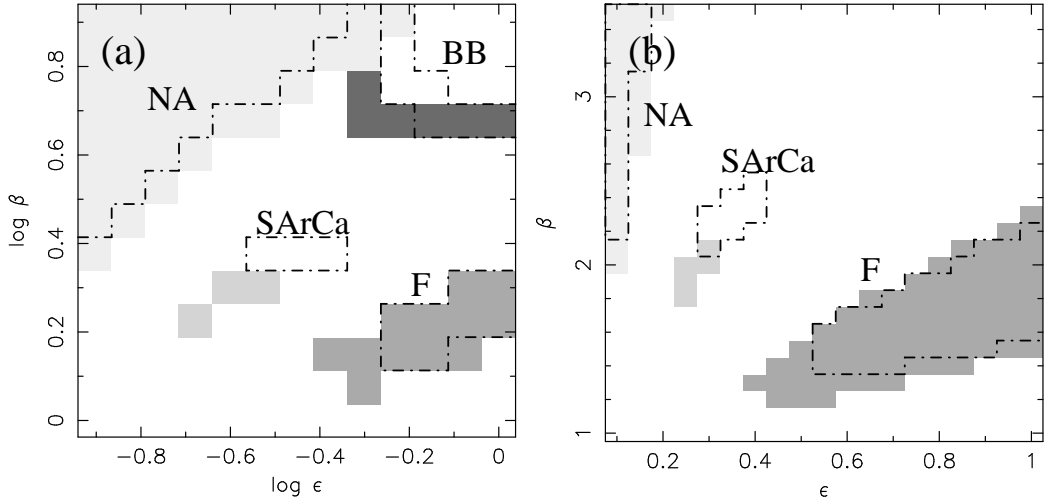}
  \caption{(a) A comparison between the region of allowed $\epsilon:\beta$
  parameter space for M82 predicted by the scaled CC model (shaded regions) in 
  comparison to the regions predicted by the full two-dimensional 
  hydrodynamical simulations (regions enclosed by dot-dashed lines, previously
  shown in Fig.~\ref{fig:new_medium_vh1}), using parameter set A.
  (b) As in (a), except the comparison uses the finer grid of models 
  distributed linearly in a smaller region of $\epsilon:\beta$ parameter 
  space.}
  \label{fig:vh1_vs_scaledcc}
\end{figure*}

As shown in Fig.~\ref{fig:vh1_vs_scaledcc}, 
these same general regions of acceptable parameter space are reproduced
in simulations making use of the scaled CC model, although in detail there
are minor difference caused by the $\la 30$\% errors in the luminosity
predicted by the scaled CC model compared to the multi-dimensional 
hydrodynamical simulations (Appendix~\ref{app:non_spherical}).

\subsection{Other input model parameters.}
\label{sec:model:other_parameters}

Given the computational cost of the full hydrodynamical
simulations it is preferable to use the scaled CC model to explore the effect
of varying other model parameters such as starburst region size, metal
abundance and participation factor. Our experience with parameter
set A indicates that the scaled
CC model reproduces the allowed values of $\epsilon$ and $\beta$
to an accuracy of $\la 0.1$ for the 
full set of observational constraints that include the 
Fe lines.

\begin{figure*}
\epsscale{0.9}
\plotone{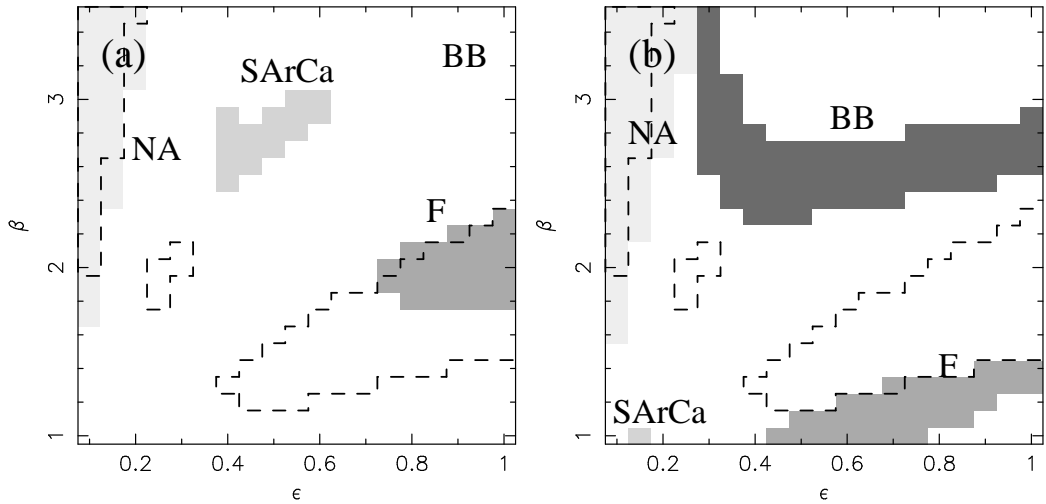}
  \caption{(a) A comparison between the region of allowed $\epsilon:\beta$
  parameter space for M82 predicted by the scaled CC model using
  parameter set B (shaded regions) in 
  comparison to the equivalent regions for parameter set A (dashed lines).
  (b) As in (a), except the comparison is between parameter set C
  and parameter set A. Parameter sets B and C differ from A in the assumed
  starburst star formation mode and age.}
  \label{fig:matches_modb_modc}
\end{figure*}

\begin{figure*}
\epsscale{0.9}
\plotone{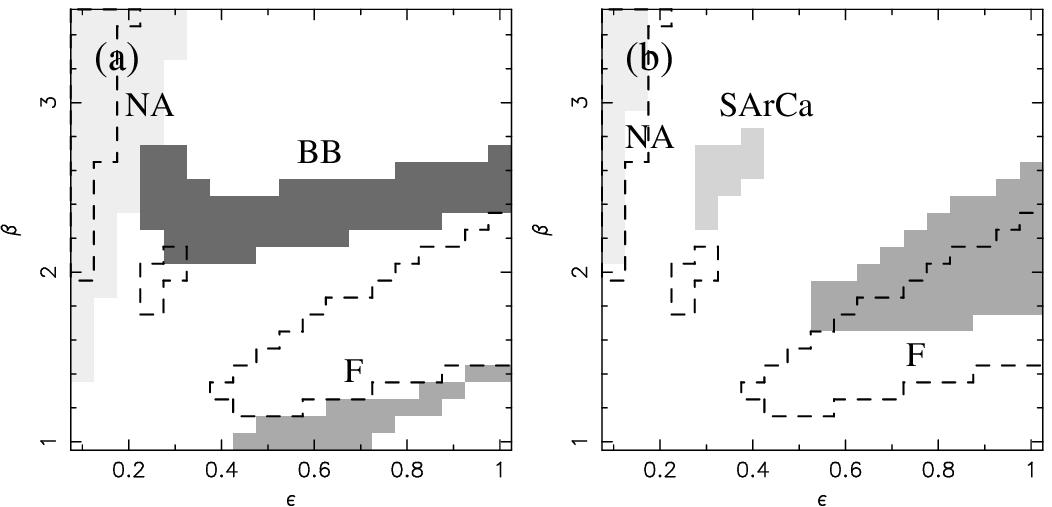}
  \caption{(a) A comparison between the region of allowed $\epsilon:\beta$
  parameter space for M82 predicted by the scaled CC model using
  parameter set D (shaded regions) in 
  comparison to the equivalent regions for parameter set A (dashed lines).
  (b) As in (a), except the comparison is between parameter set E
  and parameter set A. Parameter sets D and E differ from A in adopting
  respectively smaller and larger starburst effective radii $R_{\star}$.}
  \label{fig:matches_modd_mode}
\end{figure*}

\begin{figure*}
\epsscale{0.9}
\plotone{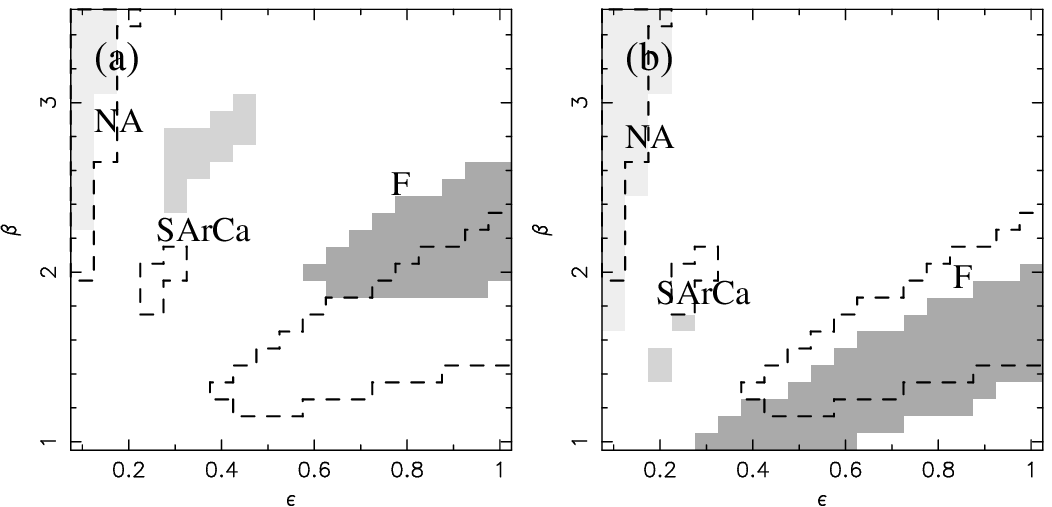}
  \caption{(a) A comparison between the region of allowed $\epsilon:\beta$
  parameter space for M82 predicted by the scaled CC model using
  parameter set F (shaded regions) in 
  comparison to the equivalent regions for parameter set A (dashed lines).
  (b) As in (a), except the comparison is between parameter set G
  and parameter set A. Parameter sets F and G differ from A in adopting
  respectively smaller and larger supernova and stellar wind metal
  abundances $Z_{\rm SN+SW}$.}
  \label{fig:matches_modf_modg}
\end{figure*}

\begin{figure*}
\epsscale{0.9}
\plotone{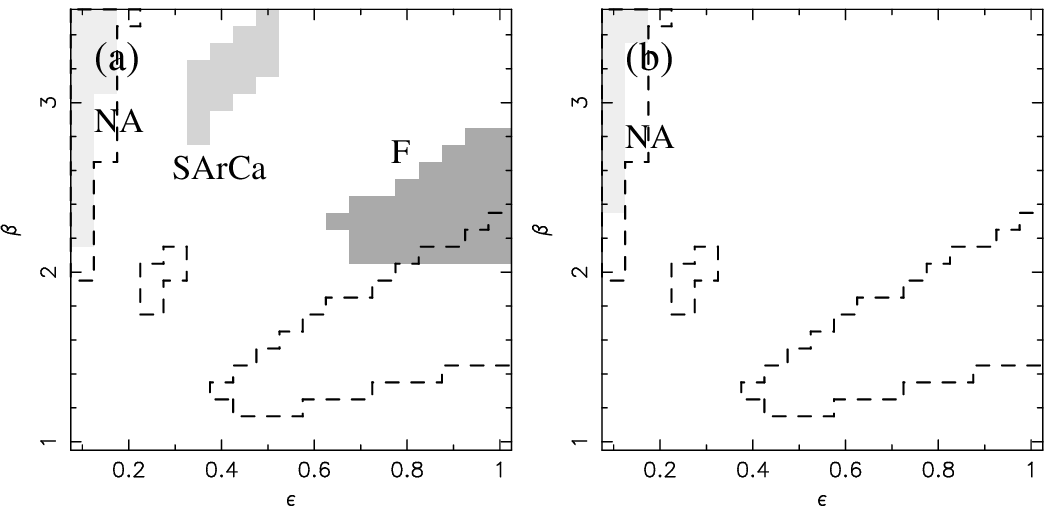}
  \caption{(a) A comparison between the region of allowed $\epsilon:\beta$
  parameter space for M82 predicted by the scaled CC model using
  parameter set H (shaded regions) in 
  comparison to the equivalent regions for parameter set A (dashed lines).
  (b) As in (a), except the comparison is between parameter set I
  and parameter set A. Parameter sets H and I differ from A in adopting
  lower values of the participation factor $\zeta = 0.71$ and $\zeta = 0.50$ 
  respectively, \ie effectively lower star formation rates.}
  \label{fig:matches_modh_modi}
\end{figure*}

Parameter sets B through I are shown in Table~\ref{tab:parameter_sets}.
These alternate parameter sets 
explore the influence of the chosen star formation mode and age,
starburst region size, supernova and stellar wind metal abundance,
and participation factor on the region of allowed $\epsilon:\beta$
parameter space. All of these calculations were performed using the
scaled CC model alone. Predicted central density, pressure and
temperature, along with the terminal velocity, total cooling, broad-band
hard X-ray ($E=2$ -- 8 keV band) luminosity and Fe He$\alpha$
and S Ly$\alpha$ line luminosities for $\epsilon = 1$ and $\beta = 1$
are shown in Table~\ref{tab:general_properties}.

\begin{deluxetable*}{lrcccccc}
 \tabletypesize{\scriptsize}%
\tablecolumns{8}
\tablewidth{0pc}
\tablecaption{Superwind properties that match the full set of observational
        constraints.
        \label{tab:allowed_regions}}
\tablehead{
\colhead{Parameter Set} 
    & \colhead{$N_{\rm match}$}
    & \multicolumn{6}{c}{Range in parameter consistent with full constraints} \\
\colhead{\nodata} 
     & \colhead{\nodata}
     & \colhead{$\epsilon$}
     & \colhead{$\beta$}
     & \colhead{$\log T_{\rm  c}$} 
     & \colhead{$\log P_{\rm c}/k$} 
     & \colhead{$\log \hbox{$\dot p$}_{\rm  WF}$} 
     & \colhead{$v_{\infty}$}
     \\
\colhead{(1)}
     & \colhead{(2)} & \colhead{(3)}
     & \colhead{(4)} & \colhead{(5)}
     & \colhead{(6)} & \colhead{(7)}
     & \colhead{(8)}
}
\startdata
A\tablenotemark{a} & 52
     & 0.55 -- 1.0  & 1.4 -- 2.2 
     & 7.56 -- 7.82 & 7.15 -- 7.38 
     & 34.31 -- 34.54 & 1555 -- 2097 \\
A    & 75 
     & 0.40 -- 1.0 & 1.2 -- 2.3 
     & 7.51 -- 7.84 & 7.08 -- 7.40 
     & 34.23 -- 34.55 & 1470 -- 2164 \\
B & 26 
     & 0.75 -- 1.0 & 1.8 -- 2.3 
     & 7.49 -- 7.66 & 7.13 -- 7.24 
     & 34.29 -- 34.39 & 1434 -- 1745 \\
C & 32 
     & 0.45 -- 1.0 & 1.0 -- 1.4 
     & 7.64 -- 7.87 & 7.22 -- 7.47 
     & 34.37 -- 34.62 & 1709 -- 2239 \\
D & 21 
     & 0.45 -- 1.0 & 1.0 -- 1.4 
     & 7.67 -- 7.87 & 7.80 -- 8.05 
     & 34.20 -- 34.44 & 1778 -- 2240 \\
E & 60 
     & 0.55 -- 1.0 & 1.7 -- 2.6 
     & 7.48 -- 7.76 & 6.94 -- 7.17 
     & 34.35 -- 34.58 & 1426 -- 1976 \\
F & 47 
     & 0.60 -- 1.0 & 1.9 -- 2.6 
     & 7.50 -- 7.72 & 7.26 -- 7.43 
     & 34.41 -- 34.58 & 1452 -- 1874 \\
G & 76 
     & 0.30 -- 1.0 & 1.0 -- 2.0 
     & 7.50 -- 7.87 & 6.96 -- 7.37 
     & 34.11 -- 34.52 & 1452 -- 2240 \\
H & 43 
     & 0.65 -- 1.0 & 2.1 -- 2.8 
     & 7.47 -- 7.70 & 7.15 -- 7.29 
     & 34.30 -- 34.44 & 1409 -- 1829 \\
I & 0 
     & \nodata & \nodata
     & \nodata & \nodata
     & \nodata & \nodata \\
\enddata
\tablenotetext{a}{Calculated using 2-dimensional hydrodynamical simulations.
  All other parameter sets were calculated using the scaled 1-dimensional
  \citeauthor{chevclegg} model.}
\tablecomments{Column 1: Parameter set name.
  Column 2: $N_{\rm match}$ is the number of models that match the full
  set of observational constraints, out of the 494 models calculated for
  each parameter set.
  Columns 3 and 4: The range of thermalization efficiencies $\epsilon$
  and mass loading factors $\beta$ found to match the full set 
  of observational constraints.
  The remaining columns show the range in a variety of physical
  properties of the M82 superwind found in the models that match the
  full set of observational constraints:
  central temperature $T_{\rm c}$ (column 5, in Kelvin),
  central pressure $P_{\rm c}/k$ (column 6, in units of K cm$^{-3}$),
  rate of change of superwind momentum 
  $\hbox{$\dot p$}_{\rm  WF} = \Mdot_{\rm tot} \times v_{\infty}$ 
  (column 7, units g cm s$^{-2}$) and wind terminal velocity 
   $v_{\infty}$ (column 8, units $\kmps$).
}
\end{deluxetable*}

Parameter set B differs from the previously discussed parameter set A only
in that is assumes that ongoing star formation began 10 Myr ago, rather than
30 Myr ago. As the younger stellar population is intrinsically more luminous,
less star formation is required to match the IR luminosity of M82 and
consequently the mechanical energy and mass injection rates in parameter
set B are lower than in parameter set A. Furthermore the ratio of
energy to mass injection is lower, resulting in a lower central
temperature of $T_{\rm c} \sim 8 \times 10^{7}$ K for unit $\epsilon$
and $\beta$ compared to $T_{\rm c} \sim 10^{8}$ K in parameter set A,
and a lower temperature than parameter set A for any given value
of  $\epsilon$ and $\beta$. Cumulatively
these differences result in a significant shift in the location of the
the models in $\epsilon:\beta$ parameter space that match either the
full or SArCa constraints (see Fig.~\ref{fig:matches_modb_modc}a).

Generation of plots equivalent to Fig.~\ref{fig:new_fine_vh1} (not shown)
reveals that the region matching the full set of constraints is again
determined by the intersection of the Fe He$\alpha$ and S Ly$\alpha$
observational constraints. Furthermore, at a given value of $\epsilon$
and $\beta$ the predicted broad band and (in particular) line luminosities are
lower for parameter set B than for parameter set A,
with the least difference in luminosity occurring at $\epsilon = \beta = 1$.

Application of Equation~\ref{equ:ei} indicates that parameter set B yields
emission integrals consistently $\sim 0.1$ dex lower than parameter set A
for all values of  $\epsilon$ and  $\beta$. For the Fe and S lines the
lower temperature of parameter set B when  $\epsilon \sim \beta \sim 1$
results in a higher cooling function that offsets the reduced emission 
integral because the lower temperature is closer to the peak of
the line-specific cooling functions, such that
the predicted values of $L_{\rm Fe He\alpha}$ and
$L_{\rm S Ly\alpha}$ are comparable to those of parameter set A 
(Table~\ref{tab:general_properties}). But over much of the $\epsilon:\beta$
parameter space the lower temperatures in parameter set B result in lower
line emissivities that in combination with the lower emission integral result
in lower line luminosities than in parameter set A. Ultimately
the regions in  $\epsilon:\beta$ parameter space that match the full and
SArCa constraints in parameter set B require higher mass loading and higher
thermalization efficiency than in parameter set A in order to increase the
emission integral without further decreasing the temperature. 
Increasing $\beta$ strongly increases the
net emission integral as $EI \propto \beta^{3}/\epsilon$,
although it slightly decreases the plasma temperature (which must then be
offset by increasing the thermalization efficiency). In contrast, reducing
the thermalization efficiency is counter productive as the increase in $EI$ is
counteracted by the decrease in temperature that consequently reduces
the line emissivities.

Parameter set C represents the star formation in the nucleus M82 as a
single instantaneous burst occurring 10 Myr ago. Compared to parameter
set A this results in a higher mechanical energy and mass injection rate,
although the ratio of energy to mass injection is much closer to that in
parameter set A than the lower values of parameter set B.

The higher mass injection for parameter set C leads to 
higher predicted X-ray luminosities for a given value of $\epsilon$
and $\beta$ than for parameter set A, which has a significant effect
on the regions of $\epsilon:\beta$ parameter space that match the
various constraints (Fig.~\ref{fig:matches_modb_modc}b).
The region matching the $E=2$ -- 8 keV broad band luminosity appears
at much lower $\beta$ than in parameter set A, although in terms
of absolute mass injection $\Mdot_{\rm tot}$ the location of this
region is consistent between parameter set A and parameter set C.

Similarly, lower values of $\beta$ are required to match the
full set of constraints, the allowed region extending between
$0.4 < \epsilon \le 1$ and $1 \le \beta \la 1.4$. Models with
$\epsilon = \beta = 1$ are again excluded as possible matches to the
observed properties of M82, but for different reasons than in the case of
parameter sets A and B. For parameter set C models with $\beta \sim 1$ and
$\epsilon \sim 1$ generate sufficient Fe He$\alpha$ luminosity to
match the required constraint, but these models predict a Fe 
Ly$\alpha$/He$\alpha$ line ratio greater than the upper limit of
0.34 (\ie $T_{c}$ is too high), 
and hence do not match the full set of constraints. A combination
of constraints shapes upper edge of the allowed region. In addition to
the requirement not to overproduce the S Ly$\alpha$ emission that we
encountered with parameter sets A and B, the upper edge of the
allowed region in parameter set C is also set where the 
Fe He$\alpha$ luminosity exceeds the observational limits.

Parameter set D adopts the compact starburst region discussed in
\S~\ref{sec:theory:applying:sbsize}. In terms of the effective
 radius of the starburst region parameter set D uses $R_{\star} = 125.5$ pc 
instead of $R_{\star} = 300$ pc, but is otherwise identical to parameter 
set A. Consequently the gas density within the starburst region 
($n_{\rm c} = 0.91 \pcc$) is considerably higher than in parameter set A 
($n_{\rm c} = 0.16 \pcc$) or even parameter set C ($n_{\rm c} = 0.26 \pcc$),
and results in a significant increase in the X-ray emission despite the
reduction in starburst region volume (see Equ.~\ref{equ:ei}). The regions of
$\epsilon:\beta$ parameter space over which parameter set D matches 
the various sets of constraints is very similar to that parameter
set C (Fig.~\ref{fig:matches_modd_mode}a) for the same reasons we discussed
above. The region of $\epsilon:\beta$
parameter space that matches the full set of constraints is smaller
than in parameter set C because of the greater increase in luminosity caused
by the higher gas density, and no region of 
 $\epsilon:\beta$ matches the SArCa constraints.
 It is possible that with tighter observational constraints 
on $L_{\rm Fe He\alpha}$
and $L_{\rm S Ly\alpha}$ there would be no region of $\epsilon:\beta$
parameter space that would match the full set of observational constraints.

The lower volume of the starburst region adopted in parameter set D also
leads to
higher central pressures. Models that match the full set of observational
constraints have $7.80 \le \log P_{\rm c}/k \le 8.05$, values that are 
2 -- 4 times higher than the highest independent observational estimates
of the thermal pressure in the core of M82. This leads us to suspect that
parameter set D is not a valid physical representation on the M82 starburst.

Parameter set E adopts the largest of the possible starburst region geometries
discussed in \S~\ref{sec:theory:applying:sbsize}, for which the effective
radius is $R_{\star} = 406$ pc and the resulting central number density is
$n_{\rm c} = 0.09 \pcc$. The region that matches the full set of observational
constraints now moves toward higher $\epsilon$ and higher $\beta$ with
respect to the same region in parameter set A 
(Fig.~\ref{fig:matches_modd_mode}b), for similar reasons to those we discussed
regarding parameter set B. Increased mass loading is required to make up for
the reduction in gas density due to the larger starburst, but the 
thermalization efficiency must also increase so as to maintain a central
temperature that maximizes the Fe He$\alpha$ emissivity ($7.5 \la \log 
T_{\rm c} \la 7.8$, for parameter sets A, B and E. See Table~\ref{tab:allowed_regions}). 

In \S~\ref{sec:theory:applying:z} we discussed the expected
metal abundance of the
merged SN ejecta and stellar wind ejecta. The calculations of \citet{limongi07}
predict a higher iron abundance for the SN ejecta 
($Z_{\rm Fe,SN} = 7 Z_{\rm Fe,\odot}$) than the Starburst99 calculations that
combine \citet{woosley95} SN yields with stellar wind ejecta 
($Z_{\rm Fe, SN+SW} = 4 Z_{\rm Fe,\odot}$). Furthermore the 
\citeauthor{limongi07} calculations predict that sulphur, argon and calcium 
are less enriched than iron, while Starburst99 predicts that sulphur 
is more strongly enriched than iron. In all of the calculations we have
discussed until now we adopted a uniform relative enrichment of all 
elements, $Z_{\rm SN+SW} = 5 Z_{\odot}$. Parameter sets F and G explore
the influence of the adopted metal abundance on the region of
$\epsilon:\beta$ parameter space that match the observational constraints:
parameter set F adopts  $Z_{\rm SN+SW} = 2.5 Z_{\odot}$ while
parameter set G adopts $Z_{\rm SN+SW} = 7.5 Z_{\odot}$. Given the
already large uncertainties in the yields we feel that exploring the
effect of adopting different yields for different elements is beyond the
scope of this paper.

The lower metal abundance in parameter set F leads to lower
lower S, Ar, Ca and Fe line luminosities than for parameter set A.
For $\beta = 1$ the line luminosities are exactly half those in
parameter set A, but the fractional difference becomes less 
as $\beta$ increases and the influence of the mass-loaded material
on the net metal abundance of the wind fluid $Z_{\rm WF}$ 
becomes larger (Equ.~\ref{equ:zwf}). 
The result in terms of the region of $\epsilon:\beta$ parameter space that 
matches the observational constraints is shown in 
Fig.~\ref{fig:matches_modf_modg}a, and is very similar to that in
parameter sets B and E that also generate lower line luminosities for
different reasons.

Conversely the higher metal abundance in parameter set G leads to higher
line luminosities than parameter set A, although they are generally less
than those 
in parameter sets C and D (see \eg Table~\ref{tab:general_properties}).
The $\epsilon:\beta$ region of matches is shown in
Fig.~\ref{fig:matches_modf_modg}b. The upper $\beta$ edge of the region
that matches the full set of observational constraints is set by the
upper limit on the S Ly$\alpha$ luminosity, while the lower  $\beta$ edge
of this region is set by a combination of the Fe Ly$\alpha$/He$\alpha$ line
ratio constraint and the minimum acceptable Fe He$\alpha$ line luminosity.

Parameter sets H and I explore the effect of varying the participation
fraction $\zeta$ (see \S~\ref{sec:theory:applying:edotmdot}). Reducing $\zeta$
has the effect of reducing the effective star formation rate without
altering the energy per particle, and the emission integral and luminosities 
for a given value of $\epsilon$ and $\beta$ are equal to those from
parameter set A except multiplied by a factor $\zeta^{2}$. For parameter
set H, where $\zeta = 2^{-1/2}$, the factor 2 reduction in all 
luminosities moves the regions that
match both the SArCa and full set of constraints to higher $\beta$
and higher $\epsilon$ (Fig.~\ref{fig:matches_modh_modi}a). 
Parameter set I has $\zeta = 0.5$ and the reduction in luminosity is
so significant that the region matching the SArCa constraints has moved
off of the region shown in
 Fig.~\ref{fig:matches_modh_modi}b to $0.5 \la \epsilon \la 0.9$,
$4.5 \la \beta \la 5.5$.
More significantly, there are no values of $\epsilon$ or $\beta$ 
that can match the full set of observational constraints when using
parameter set I.

\begin{deluxetable*}{lllcccrrrr}
 \tabletypesize{\scriptsize}%
\tablecolumns{10}
\tablewidth{0pc}
\tablecaption{Starburst region properties for $\epsilon = 1$ and
  $\beta = 1$.
        \label{tab:general_properties}}
\tablehead{
\colhead{Parameter Set}
     & \colhead{$\log n_{\rm c}$}
     & \colhead{$\log P_{\rm c}/k$} 
     & \colhead{$\log T_{\rm  c}$} 
     & \colhead{$v_{\infty}$}
     & \colhead{$r_{12}$} 
     & \colhead{$\log L_{\rm tot}$} 
     & \colhead{$\log L_{\rm 2-8 keV}$} 
     & \colhead{$\log L_{\rm Fe He\alpha}$} 
     & \colhead{$\log L_{\rm S Ly\alpha}$} 
     \\
\colhead{\nodata}
     & \colhead{\nodata} 
     & \colhead{\nodata} 
     & \colhead{\nodata} 
     & \colhead{($\kmps$)}
     & \colhead{(pc)} 
     & \colhead{\nodata}
     & \colhead{\nodata}
     & \colhead{\nodata}
     & \colhead{\nodata}
     \\
\colhead{(1)}
     & \colhead{(2)} & \colhead{(3)}
     & \colhead{(4)} & \colhead{(5)}
     & \colhead{(6)} & \colhead{(7)}
     & \colhead{(8)} & \colhead{(9)}
     & \colhead{(10)}
}
\startdata
A   &   -0.80  &    7.22  &  8.02
  &    2651  &   278  &     38.67  &     38.27  &     37.46  &     36.22  \\ 
B   &   -0.85  &    7.06  &  7.91
  &    2341  &   278  &     38.56  &     38.17  &     37.45  &     36.24  \\ 
C   &   -0.59  &    7.40  &  7.98
  &    2548  &   285  &     39.09  &     38.70  &     37.92  &     36.69  \\ 
D   &   -0.04  &    7.98  &  8.02 
  &    2651  &   121  &     39.17  &     38.75  &     37.93  &     36.74  \\ 
E   &   -1.06  &    6.96  &  8.02 
  &    2651  &   369  &     38.49  &     38.09  &     37.27  &     35.98  \\ 
F   &   -0.80  &    7.22  &  8.02 
  &    2651  &   278  &     38.51  &     38.11  &     37.16  &     35.92  \\ 
G   &   -0.80  &    7.22  &  8.02 
  &    2651  &   278  &     38.80  &     38.39  &     37.64  &     36.40  \\ 
H   &   -0.95  &    7.07  &  8.02 
  &    2651  &   272  &     38.37  &     37.97  &     37.16  &     35.92  \\ 
I   &   -1.10  &    6.92  &  8.02  
  &    2651  &   266  &     38.07  &     37.67  &     36.86  &     35.62  \\ 

\enddata
\tablecomments{Column 1: parameter set name.
  Columns 2, 3 and 4: Logarithms of the central number density ($\pcc$),
  the central pressure divided by Boltzmann's constant, and the central
  temperature (K) respectively for models where $\epsilon = 1$ and $\beta = 1$.
  Column 5: Wind terminal velocity. 
  Column 6: The radius at which the ionization parameter $\eta$ 
  (see Equation~\ref{equ:eta}) drops below $10^{12} \pcc \s$.
  Columns 7, 8, 9 and 10: Logarithms of the plasma luminosity ($\ergps$)
  within a radius
  of 500 pc. Columns are the total cooling luminosity, 
  the $E=2$ -- 8 keV energy band X-ray luminosity, 
  the Fe He$\alpha$ line luminosity and the S Ly$\alpha$ line luminosity
  respectively.}
\end{deluxetable*}

\section{Discussion}
\label{sec:discussion}

In \citet{strickland07} we concluded that the diffuse Fe He$\alpha$ emission
from the central $\sim 500$ pc of M82 could have been produced by thermalized
SN ejecta and stellar wind material from the starburst based on relatively
simple estimates of the plasma properties of the emitting material.

The more rigorous approach to the interpretation of the diffuse hard
X-ray emission taken in this paper strengthens that conclusion:
the plasma responsible for the hard X-ray iron-line emission 
is the same gas that drives the larger-scale superwind in
this starburst galaxy. 
Furthermore our method allows us to make quantitative statements regarding
supernova feedback and the properties of M82's starburst-driven superwind.

\subsection{Supernova Thermalization (Heating) Efficiencies and Mass Loading}
\label{sec:discussion:eta_beta}

The primary aim of this paper was to establish what constraints could
be made on the efficiency of supernova heating and the net rate of
mass injection within a starburst region.

It is generally accepted that in more normal, or quiescent, star forming
galaxies such as the Milky Way on average only $\sim 10$\% of the $10^{51}$ 
erg of initial kinetic energy per supernova ends up heating or moving the
ISM, with the remainder being lost as radiation 
\citep[see \eg][]{thornton98,efstathiou00}. Although it had long been 
suspected that the efficiency of supernova heating is a function of the
local interstellar and star forming conditions there was no direct
method by which to measure this thermalization efficiency in a starburst
galaxy until, as described in \S~\ref{sec:introduction}, 
the launch of the {\it Chandra} X-ray Observatory. 

Theoretical arguments have been made
for either lower or higher SN heating efficiencies in more actively
star-forming galaxies. 
\citet{steinmetz99} argued that radiative energy losses must always be
severe because massive stars form in regions of high gas density. In contrast
\citet{ham90} argue that thermalization must be efficient in starbursts 
because a high supernova rate per unit volume leads to an increased filling
factor of the hot low density ISM, which in turn reduces the radiative energy
losses experienced by subsequent SNe. Thus the SN thermalization 
efficiency depends on the SN-modulated porosity of the ISM 
\citep[\eg][]{mo77}. Recent numerical simulations by \citet{melioli04}
appear to qualitatively demonstrate such a time-dependent non-linear 
thermalization efficiency. 
Intriguingly they find that the thermalization efficiency (which they
term the heating efficiency) can only be low ($\la 10$\%) or very
high ($\sim 100$\%) for any significant 
period of time, transitioning between to two states very rapidly.

Let us briefly review how the hard X-ray line emission from a starburst
region is related the average thermalization efficiency $\epsilon$ and 
the mass loading factor $\beta$. Both of these parameters
strongly affect the net volume emission integral $EI \propto 
\beta^{3}/\epsilon$ (Equation~\ref{equ:ei}) and the line emissivities 
$\Lambda(T,Z)$ through the plasma temperature $T \propto \epsilon/\beta$ 
(Equation~\ref{equ:final_tc}, Fig.~\ref{fig:emis}) and the
metal abundance of the wind fluid $Z_{\rm WF}$ (Equation~\ref{equ:zwf}).

In the specific case of M82 it is the Fe He$\alpha$ and S Ly$\alpha$ lines 
that drive the allowed
values of $\epsilon$ and $\beta$ we derive. For many of the
sets of model parameters we have explored it should be noted that 
$\epsilon = \beta = 1$ will not reproduce the observed constraints
because too little Fe He$\alpha$ emission is produced as the plasma
density within the starburst region is too low. Thus successful models
must have a higher volume emission integral  which implies $\beta > 1$ 
and/or $\epsilon < 1$.

As we described in \S~\ref{sec:model:epsilon_and_beta},
the lower boundary of acceptable mass loading factors $\beta$
for a given value of the 
thermalization efficiency $\epsilon$
is set by requirement to achieve sufficient Fe He$\alpha$ emission
($37.96 \le \log L_{\rm Fe He\alpha} \le 38.39$),
while the maximum $\beta$ is set by the requirement not to exceed the observed
S Ly$\alpha$ luminosity ($\log L_{\rm S Ly\alpha} \le 37.43$). 

Furthermore the central
temperature (and hence $\epsilon/\beta$)
must be high enough that the Fe He$\alpha$ emissivity is
significant with respect to the emissivity of the lower ionization lines,
in particular the sulphur lines. This constraint ultimately controls 
the minimum 
allowed value of the thermalization efficiency $\epsilon$, and is the
reason why the different parameter sets tend to have similar minimum
values of $T_{\rm c}$, specifically $\log T_{\rm c} \ga 7.5$ 
(see Table.~\ref{tab:allowed_regions}).

The central temperature can not exceed $7.8 \times 10^{7}$ K 
($\log T_{\rm c} = 7.9$)
 as otherwise the upper limit on 
the Fe Ly$\alpha$/He$\alpha$ line ratio would be violated. For a given
value of $\epsilon$ this constraint determines the lowest allowed
values of $\beta$ in cases where the Fe He$\alpha$ luminosity is
sufficiently high enough to match the observational constraints.
This upper temperature 
constraint is only significant when the model parameters generate higher
central densities or luminosities (more mass injection, a more compact
starburst, or a higher metal abundance in the ejecta), and thus
there is a broader range in the maximum allowed $T_{\rm c}$ than
in the minimum allowed  $T_{\rm c}$.

Returning to the consideration of our results, we find that 
for all parameter sets that produce at least one
valid solution that can match the
full set of observational constraints for M82, those valid solutions
always include thermalization efficiencies of 100\%.

Furthermore $\epsilon$ is never less that 30\%, and over the
range of parameter sets we have explored the minimum valid thermalization
efficiency $\epsilon_{\rm min}$ ranges from 30\% to 75\% (see 
Table~\ref{tab:allowed_regions}). Within this range there is
a small degree of variation associated with different 
starburst model parameters
we have explored.
We find that the starburst history and ejecta metal abundance
$Z_{\rm SN+SW}$ have some effect on the allowed thermalization efficiency,
while the starburst region size appears to have little influence on the minimum
allowed thermalization efficiency (Table~\ref{tab:allowed_regions}). 
 For example in a younger continuous star formation 
(CSF) model for the burst (parameter set B) the minimum allowed 
thermalization efficiency is $\epsilon_{\rm min} = 0.75$ 
in comparison to the older CSF we adopt as the default starburst history
($\epsilon_{\rm min} = 0.4$) or the single instantaneous
burst model (parameter set C, $\epsilon_{\rm min} = 0.45$).

That $\epsilon$ must be $\ga 30$\% is a striking result. 
If the diffuse $E\sim 6.7$ keV 
iron line in M82 arises in
a collisional plasma then the average thermalization efficiency in the
M82 starburst region can be as high as 100\%, while it can not be
as low as 10\%. Our results give quantitative support to the arguments
presented by \citet{ham90} and \citet{melioli04}
for a higher thermalization efficiency in starbursts than in normal
star-forming galaxies. The issue of whether the thermalization efficiency
must be 100\%, or whether high but intermediate values are allowed, remains
an interesting but open question.

With respect to the mass-loading of the superwind in M82 we find that
the majority of the parameter sets we have considered require some
centralized mass loading of the wind fluid, although at low-to-moderate levels. Over all
parameter sets $1 \le \beta \le 2.8$, but only parameter sets C, D and
G allow $\beta = 1$ and more typically $1.5 \la \beta \la 2.5$.

Parameter sets C, D and G 
allow zero additional mass-loading ($\beta = 1$) as a valid
solution when coupled with lower values 
of the thermalization efficiency ($\epsilon \sim 30$ -- 70\%). In
parameter set C this is because the single instantaneous starburst
injects more mass than the constant star formation model used in the
other parameter sets, and hence does not require additional sources of
mass to increase the plasma density within the starburst region. 
In parameter sets D and G the non-mass-loaded plasma within 
the starburst
region is sufficiently luminous without additional mass-loading because
of either the compactness of the starburst (D), or the higher metal abundance 
(G), increases the net luminosity. High values
of $\epsilon$ are excluded as valid solutions when $\beta=1$
for these three parameter sets because the
high central plasma temperature violates the Fe Ly$\alpha$/He$\alpha$
line ratio constraints. 

Thus these results suggest that it is plausible that
the raw SN and stellar wind ejecta
from the massive stars in M82's starburst have been efficiently mixed with
a roughly equivalent mass of initially cooler ambient gas, although
no mass loading is also a possible solution. The net effect
is an average mass outflow rate from the starburst region
in the very hot phase of the wind fluid 
of $\Mdot_{\rm tot} \sim 2.5 \Msol \pyr$, approximately half of the gas
consumption due to the star formation rate alone 
(\S~\ref{sec:theory:applying:sbmag}). The allowed range of wind fluid
outflow rate covering all parameter sets except parameter set D 
lies in the range $1.4 \la \Mdot_{\rm tot} (\Msol \pyr)
\la 3.6$ (a fraction $\sim 0.3$ -- 0.8 of the star formation rate
for the Salpeter IMF we adopt).

This level of mass-loading is significantly lower than the values others
have derived from soft X-ray observations of superwinds. \citet{suchkov96}
used  fairly sophisticated 1-dimensional hydrodynamical modeling to estimate
that a mass-loading factor
of $\beta \sim 3$ -- 6 was required in to match the soft
X-ray properties of M82 known at that time from {\it Einstein} and 
{\it ROSAT} HRI observations. For the dwarf
starburst galaxy NGC 1569 \citet{martin02} estimated 
$\beta \sim (8$ -- $35) \times \eta^{1/2}$ (where $\eta$ is the volume
filling factor of the X-ray-emitting plasma) based on the estimated
mass of soft X-ray-emitting gas seen in {\it Chandra} 
observations. \citet{grimes07} derived a value of $\beta \sim 5 \, \eta^{1/2}$
for the soft X-ray-emitting plasma in the local Lyman Break Galaxy analog
Haro 11.
Both \citeauthor{martin02} and \citeauthor{grimes07} obtain
similar estimates of $\beta$ based on the best-fit metal abundances 
in the soft X-ray-emitting plasma compared to the expected abundances 
of a SN-ejecta-enriched plasma (see Equation~\ref{equ:zwf}).

However a direct comparison between our findings and the existing results
in the literature is somewhat misleading.

The results from the literature
are all based on soft X-ray observations of starbursts that probe
gas in the temperature range 2 -- 10 million K, not the hotter
30 -- 70 million gas probed by the observations of M82's diffuse hard X-ray
emission. 
The high mass loading factors simply 
reflect the fact that there is more mass in the
soft X-ray emitting plasma than can be accounted for by SN and stellar
wind ejecta alone. 

Soft X-ray mass loading factors are further
based on measurements that cover the entire
soft X-ray nebula associated with the target starburst galaxy 
(over $\sim 5$ kpc-scales for M82 and Haro 11, and over 
$\sim 1$ kpc in NGC 1569), whereas our measurement constrains 
the mass loading of the wind fluid
within the $\sim 500$ pc starburst region 
itself (in the terminology introduced by
\citet{suchkov96} this is central mass loading). It is plausible that
the wind fluid experiences additional mass-loading as it flows
outward beyond the starburst region (distributed mass loading).

The older measurements referred to above can not differentiate
between a localized or a global mass-loading process (by global we mean
a process that is essentially volume filling), although we
have previously argued that a significant fraction of the soft
X-ray emission must arise from a low-volume filling factor component of the
wind \citep{strickland00,strickland02,strickland04a}. In contrast 
the mass loading factor derived in this paper is for a volume-filling
mass loading.

It is worth noting that it is not the case 
that the hottest plasma in a superwind is
necessarily also the least mass-loaded phase.
The technique we have presented could detect a mass loading factor
of $\sim 5$ -- 10 for the very hot gas in the starburst region. 
The line luminosities are strong functions of the mass loading factor 
(Fig.~\ref{fig:new_medium_vh1}), so that both the Fe and S line emission
would be more intense than observed in the case of a higher mass-loading
factor. Both low thermalization efficiency and high mass loading are 
more readily observable.
In a sense it is the relative faintness of these lines from M82 that
force us toward the allowed values of the thermalization
efficiency and mass loading factor given in Table~\ref{tab:allowed_regions}.
Even if part of the starburst region experienced higher mass loading than
another part our final estimate of the allowed mass loading factor
would be skewed toward the higher
value as $\beta$ is essentially luminosity-weighted.
Thus we conclude that the $T \ga 3 \times 10^{7}$ K component of M82's
superwind (the wind fluid) genuinely has a relatively low degree 
of mass-loading within
the starburst region.

\subsection{Physical Properties of the M82 Starburst}
\label{sec:discussion:winds}

Despite the broad range of input model parameters we have explored in the
various parameter sets we find that the physical properties of the base
of the M82 superwind in the allowed models are all relatively similar 
(Table~\ref{tab:allowed_regions}).

We have already described the relatively narrow allowed range for the
central temperature, $7.5 \la \log T_{\rm c} \la 7.9$, and the physical
constraints that drive this result. The result is all the more notable given
the relatively poor constraints provided by the traditional line-ratio
diagnostics shown in Table~\ref{tab:temp_estimates}.

Similarly the central pressure and wind momentum injection rate
are constrained to lie within a range of 
$\sim 0.5$ dex ($7.0 \la \log P_{\rm c}/k \la 7.5$,
$34.1 \la \log {\dot p}_{\rm WF} \la 34.6$), if we exclude
parameter set D from consideration 
because of its unrealistically high pressure. These pressures are in good
agreement with independent estimates of the thermal pressure within
the ISM in the starburst region, as previously discussed. 

The
momentum injection rate is an important parameter with regard to 
the acceleration of the WNM and WIM by the wind fluid 
\citep[see \eg][]{ham90,heckman2000}\footnote{The terminal
velocity of a cloud of radius and mass $R_{\rm cloud}$ and $M_{\rm cloud}$,
initially exposed to a radially-expanding SN-driven wind at a radius 
$R_{\rm i}$ from the origin, is proportional to 
$({\dot p}_{\rm WF}/R_{i} \times R_{\rm cloud}^{2}/M_{\rm cloud})^{1/2}$. 
Notably the terminal velocity of the cloud can be lower than the terminal
velocity of the wind.
For a radiation-pressure driven cloud the final result is essentially the same,
with the substitution of ${\dot p}_{\rm rad} = L_{\rm BOL}/c$ 
for ${\dot p}_{\rm WF}$. Clouds of entrained material 
are effectively momentum-driven irrespective
of whether they are in an energy-driven hot wind or a radiation-driven wind.}. 
By way of comparison
the momentum injection rate
assuming 100\% thermalization efficiency and no mass loading is
$\log {\dot p}_{\rm WF} = 34.2$ for the star formation models adopted
in parameter sets A and B, and $\log {\dot p}_{\rm WF} = 34.6$ for parameter
set C. Thus taking account of imperfect thermalization and additional
(central) mass loading does not significantly alter the M82 starburst
momentum injection rate from the values one would estimate from
SNe and stellar winds alone.

\subsection{Characteristic Velocities of the M82 Superwind}
\label{sec:discussion:velocity}

The moderately tight limits on the central temperature we have
derived also make for tight limits on the terminal velocity of the wind
fluid, as $v_{\infty} \propto T_{\rm c}^{1/2}$ 
(Equation~\ref{equ:vinf}). Over all the parameter sets, 
models that match the full set of the observational constraints
predict $v_{\infty}$ in the range 1410 -- 2240 $\kmps$.

This estimate of $v_{\infty}$ is significantly higher than previous 
estimates of the terminal velocities in local starbursts (including M82) 
based on the temperature of the soft X-ray emitting gas,
$v_{\rm \infty, SX} = 440 \times (kT_{\rm SX}/0.25 \keV)^{1/2} \kmps$
\citep[see \eg][]{martin99,heckman2000,martin05}. However, it has been
long understood that these estimates were effectively lower limits
on $v_{\infty}$ given the technical difficulties associated with
robustly detecting a hotter and more tenuous (and thus intrinsically
fainter) component of the wind.

Strictly  speaking the terminal velocity is the characteristic
velocity the wind fluid
would asymptote to once outside the starburst region, if it expanded
into a vacuum (see Fig.~\ref{fig:radial_solution}). For a superwind expanding
into a gaseous medium there are additional characteristic velocities,
some of which are phase-dependent 
\citep[Strickland \& Dinge, in preparation]{ss2000}, but all
are less than or equivalent to $v_{\infty}$. Nevertheless $v_{\infty}$ remains
an accurate measure of both the effective energy per particle of the
most metal-enriched gas and the maximum velocity attainable by gas in
a pressure-driven superwind. 

For a wind that has broken-out of both the gaseous
disk and halo of its host galaxy a comparison between $v_{\infty}$ and the
local escape velocity from the gravitational potential $v_{\rm esc}$ should
give a qualitative indication of whether the material in the wind
can escape. 

Estimates of M82's escape velocity (and mass) 
have surprisingly large uncertainties for such a well-studied galaxy,
because of its very unusual rotation curve and the possibility that
its encounter with M81 has stripped it of its dark matter
halo \citep{sofue98}. The extreme options are a potential dominated
by either a central mass within a radius of $\sim 2$ kpc 
or a dark matter halo extending out to 100 kpc, the two of which crudely
limit the escape velocity to lie in the range $v_{\rm esc} \sim (1.5$
-- $3.5) \times v_{\rm rot}$. For a peak rotational velocity of
$v_{\rm rot} \sim 130 \kmps$ at a radius of $\sim 500$ pc \citep{gotz90},
the escape velocity is $v_{\rm esc} \sim 200$ -- $460 \kmps$, much lower
than the allowed range of $v_{\infty} = 1410$ -- $2240 \kmps$.

%
%
%
%
%

Our estimated terminal velocity for M82's wind
is also much larger than estimates of the
velocity of the \halpha~emitting warm ionized medium,
which achieves a maximum inclination-corrected velocity
of $v_{\rm H\alpha} \sim 600$ -- $650 \kmps$ at heights 
between $|z| \sim 350$ to 2000 pc from  
the nucleus \citep{mckeith95,shopbell98}. 
This is not surprising, as the \citet{chevclegg} model predicts a large
difference between the velocities of the wind fluid and
the velocity of the WIM (which they identified as clouds of cooler denser
gas entrained and accelerated by the wind fluid). 
In the case of M82
velocity of the WIM component of the wind appears to exceed 
the galactic escape velocity,
although more generally the WNM and WIM in superwinds 
has a speed more closely comparable to $v_{\rm esc}$ and hence
its long term fate is ambiguous \citep{heckman2000,martin05}.

Our results constrain the energy per particle within the starburst region,
neglecting the possibility of radiative energy losses and additional
mass-loading of the wind fluid exterior to the starburst region. Is it
plausible that these effects could be of sufficient magnitude
to reduce the terminal velocity of the wind fluid at large radius to
a value below the escape velocity? For a starburst with a SN thermalization
efficiency of $\epsilon=0.3$ and a central mass loading factor $\beta=1$
the larger scale wind would have to lose $\sim 90$\% of its remaining
energy to reduce the final value of $v_{\infty}$ to $460 \kmps$
in the absence of additional mass loading of the wind fluid
($98$\% energy loss to reduce $v_{\infty}$ to $200 \kmps$). The
required energy loss fraction is very similar for a central starburst with
$\epsilon=1$ and $\beta=2.3$.
We know that radiative energy losses in M82's large-scale 
wind are small, $\la 10$\% \citep{hoopes03,strickland04b}. It
seems unlikely that the wind fluid could so effective in transferring enough
of its energy to a different gaseous phase, such as the galactic halo medium,
without a significant fraction of this energy being radiated and hence
observed, unless the radiation is spread over a region much larger than
the currently studied $\sim 12$ kpc extent of M82's wind. 
The remaining alternative is to strongly mass load the wind fluid
once it has left the starburst region. Assuming that energy losses in the
wind fluid are negligible then in the case of the starburst with
$\epsilon=0.3$ and $\beta=1$ we would need to effectively mix
an additional $\Mdot_{\rm extra} \sim 13 \Msol \pyr$ into the wind
fluid outside the starburst region
to reduce $v_{\infty}$ to $460 \kmps$
($\sim 70 \Msol \pyr$ to reduce $v_{\infty}$ to $200 \kmps$).
For $\epsilon=1$ and $\beta=2.3$ then $\Mdot_{\rm extra} \sim 43 \Msol \pyr$
($v_{\infty} = 460 \kmps$) and $ \sim 240 \Msol \pyr$
($v_{\infty} = 200 \kmps$). Even the smallest of these values is uncomfortably
large in terms of the required gas reservoir, and furthermore the
resulting X-ray emissivity of the volume filling wind 
fluid might then exceed the observed soft X-ray luminosity 
of the wind ($EI \propto \Mdot^{3}$). 
Given the preceding  arguments
we believe that is probable that the
metal enriched wind fluid will ultimately escape M82.

\subsection{Implications for Understanding Feedback and Superwinds}
\label{sec:discussion:implications}

It is worth stressing that a multi-phase wind model leads to phase-dependent
ejection efficiencies for the material in starburst-driven superwinds. 
Current cosmological models of galaxies that
attempt to include superwinds 
invariably have single phase winds
(although the simulation may allow multiple ISM phases, their
galactic winds lack the fine-scale multiphase nature of real winds).
Cosmological simulations lack the ability to separately model 
the differential
ejection of the metal-enriched high-energy-per-particle wind fluid
from the less enriched, lower velocity, but more massive WNM and WIM.
Their results on metal transport and feedback processes in galaxies 
are potentially misleading or incorrect.

In particular, simulations that add all the gas mass of the WNM and WIM
phases to the metal-enriched SN ejecta 
require nonphysically large fractions ($\epsilon 
\ga 100$\%) of all the available SN
mechanical energy in star-forming galaxies (normal galaxies in addition
to starburst galaxies) in order
to match the observed degree of metal loss from galaxies and metal
enrichment of the IGM.
 \citep[\eg][]{springel03b,oppenheimer06,kobayashi07}.
These papers interpret such problems as implying the physical 
inadequacy of purely SN-driven winds and need for addition energy or wind 
sources (\eg radiation-driven winds). Our interpretation is that
the flaw lies in the inadequacy of their recipe for winds. 
In contrast simulations that
effectively treat the SN ejecta alone without additional WIM/WNM mass loading
achieve good matches to observed Ly$\alpha$ and \ion{O}{6} absorber
statistics while using reasonable SN efficiencies 
\citep[$\epsilon \sim 30$\%,][]{cen05,cen06a,cen06b}.

We further note that it is unrealistic to apply a single uniform feedback 
prescription to all star forming galaxies. 
As discussed previously there are good reasons to expect
that $\epsilon$ is a function of the local intensity of 
star formation. Furthermore 
only galaxies with average star formation rates\footnote{This value for
the critical $\Sigma_{\rm SF}$ is based on
a Salpeter IMF with mass limits of 1 and 100 $\Msol$. \citet{dahlem06} 
demonstrate that the star formation intensity 
threshold to establish extra-planar gaseous activity
depends on the host galaxy mass, but as this mass dependence is relatively
weak we have ignored it in presenting the simple superwind recipe.} 
per unit area 
$\Sigma_{\rm SF} \ga  4 \times 10^{-2} \Msol \pyr$ kpc$^{-2}$
show evidence for superwinds \citep{lehnert96b,heckman03,dahlem06}. 

In the spirit of \citet{martin99} we now present a simplified bimodal
feedback and superwind recipe appropriate for use in semi-analytical 
and numerical simulations of galaxy formation. This recipe is most
applicable to simulations
that aim to answer questions regarding the galaxy mass-metallicity
relationship, and/or the metal content of the IGM, as it primarily 
addresses the
metal-enriched stellar wind and SN ejecta from starburst-driven
superwinds. 

The formulation presented below differs
from most currently used recipes for winds.  
It prevents massive galaxies with moderate star 
formation rates (such as the Milky-Way) and the bulk of the galaxy
population at low redshift from having any wind-related 
metal-loss, but the chances of metal ejection from
intensely star-forming galaxies (starbursts) is enhanced.
It is 
unlikely to act as an effective mechanism to suppress star formation 
in all star-forming galaxies, unlike the wind recipes in the simulations we 
criticized earlier.

In galaxies with average star formation rates per units area
greater than $\Sigma_{\rm SF} \ge  4 \times 10^{-2} \Msol \pyr$ kpc$^{-2}$
(\ie starbursts) feedback is efficient and leads to the creation
of galactic superwinds.
The hot gas in a superwind can be characterized by a mass injection
rate $\Mdot_{\rm hot}$ of material of metal abundance $Z_{\rm hot}$, 
with an associated mechanical
energy injection rate $\Edot_{\rm hot}$. In terms of
the star formation rate $\cal{S}$ in units of $\Msol \pyr$ for
a Salpeter IMF between the mass limits of 1 and 100 $\Msol$, then:
\begin{itemize}
\item $\Mdot_{\rm hot} \approx 0.3 \times \beta_{\rm hot} \, \zeta \,
  \cal{S} \, \Msol \pyr,$ where $\beta_{\rm hot}$ is in the range 
  1.0 -- 2.8.
\item $Z_{\rm hot}$ is given by Equation~\ref{equ:zwf}, and is a function of
  the metal abundance in the merged SN and stellar wind eject $Z_{\rm SN+SW}$, 
  the metal abundance in the cool ambient ISM $Z_{\rm cold}$ and the 
  mass loading factor $\beta_{\rm hot}$.
\item The effective mechanical energy injection rate, 
  accounting for the efficiency of thermalization, 
  $\Edot_{\rm hot} \approx 6.5 \times 10^{41} \time \epsilon_{\rm hot}
  \, \zeta \, \cal{S} \ergps,$ where $\epsilon_{\rm hot}$ is in the range
  0.3 -- 1.0.
\item Alternatively, if winds are implemented by imparting a fixed velocity
  to gas ``particles,'' then the appropriate velocity is 
  $v_{\rm hot} \approx 2650 \, 
  (\epsilon_{\rm hot}/\beta_{\rm hot})^{1/2} \kmps$,
  and should be in the range 1400 -- $2200 \kmps$.
\item The values of $\epsilon_{\rm hot}$ and $\beta_{\rm hot}$ are correlated
  in the sense that low values of $\epsilon$ are associated with low values
  of $\beta$, such that $T_{\rm c}$ or $v_{\infty}$ lie in the range
  described in \S~\ref{sec:discussion:eta_beta} \& 
  \ref{sec:discussion:velocity}.
\item The participation factor $\zeta$ is the fraction of the total
  star formation that can effectively participate in driving a
  superwind. In the case of M82 $\zeta$ is unlikely to be lower than
  $\sim 0.5$, and is more plausible in the range $\zeta \sim 0.7$ -- 1.0
  (see Appendix~\ref{app:zeta}).
\end{itemize}

In more quiescent star forming galaxies with
 $\Sigma_{\rm SF} <  4 \times 10^{-2} \Msol \pyr$ 
kpc$^{-2}$ feedback is energetically
inefficient and no true wind arises, although SN-heated and enriched gas
may vent into the halo to form a galactic fountain. Galactic fountains
are considerably less constrained theoretically than superwinds, but it is
plausible that all the material eventually returns to the galactic disk
after some delay.

This recipe captures two essential inter-related features 
of starburst and feedback physics
that have been established in the local Universe and that have not been applied
in existing recipes for winds in cosmological simulations: the existence of
a threshold in star formation activity per unit disk area necessary to
drive global galactic winds, and that the quantitative properties of 
feedback are not the same in all galaxies but depend on the intensity of
star formation.

\subsection{Measuring feedback parameters in other starburst galaxies}
\label{sec:discussion:other_galaxies}

One limitation of the superwind recipe as presented is that it treats all
starbursts equally, and applies an M82-like wind to them all. Although
simulations of the type made by \citet{melioli04} suggest that the efficiency
of supernova feedback may be bimodal, other evidence suggests that not
all starburst-driven flows are equally 
energy-efficient \citep[see \eg][]{calzetti04}.

The techniques described in \citet{strickland07} and in
this paper can be applied to other starburst
galaxies in order to measure their effective supernova thermalization 
efficiencies and mass-loading parameters. In general one would be limited to
a purely spectral X-ray analysis, as for more distant starbursts it would
be difficult to spatially separate
the hard diffuse X-ray emission from the 
emission from X-ray binaries and nuclear AGN
even with the {\it Chandra} X-ray Observatory.

If our model is correct then we expect that the luminosity of the diffuse
hard X-ray (HX) emission of a starburst region of star formation rate $\cal{S}$
and effective radius $R_{\star}$ to scale roughly as
\begin{equation}
L_{\rm HX} \propto EI \times \Lambda
   \propto \frac{\beta^{3} \, \zeta^{2}}{\epsilon} \, \frac{{\cal{S}}^{2}}{R_{\star}},
\label{equ:lsfr_pred}
\end{equation}
where we have substituted the star formation rate into Equation~\ref{equ:ei},
and have ignored the $\epsilon$ and $\beta$ dependence of $\Lambda$.

This is notable as it predicts $L_{\rm HX}$ from the wind fluid has
a stronger than linear dependence on the star formation rate. It
should therefore be possible to extract this emission component from the
other hard X-ray emission components in star forming galaxies that
have been shown to scale linearly with star formation rate 
\citep{ranalli03,grimm03,colbert04,persic07}. The task will also be 
simplified in that it is the hard X-ray line emission that has proven itself
to provide the strongest constraints on $\epsilon$ and $\beta$ in M82,
rather than the hard X-ray continuum emission
that the X-ray binaries dominate 
in the integrated X-ray spectra of galaxies.

We note that Fe He$\alpha$ emission has been detected in the X-ray spectra
of the nearby starburst NGC 253 \citep{weaver02}, the merging system
Arp 299 \citep{ballo04}, and the ultra luminous
IR galaxies (ULIRGs) Arp 220 \citep{iwasawa05}, Mrk 273 
\citep{balestra05}, NGC 6240 \citep{netzer05}
and the Superantennae (Braito \etal 2008, submitted).
Previously this emission has been interpreted in terms of ionized reflection
associated with an AGN \citep{ross05}, and while this is likely true in 
objects with strong AGN (\eg NGC 6240), ionized reflection 
might not be the cause of the
Fe He$\alpha$ emission in all of these objects. These papers often
do consider a starburst origin for the ionized iron emission, but dismiss
it based on the expectation that its luminosity should scale linearly
with the star formation rate.

Apparently diffuse hard X-ray emission has also been seen in the {\it Chandra}
observations of the starburst galaxies NGC 2146 \citep{inui05} and NGC 3256 
\citep{lira02}, although these papers do not discuss hard X-ray line 
emission. The {\it XMM-Newton} X-ray spectrum of NGC 3256 presented 
as Figure 2 of \citet{jenkins04} shows a hint of a hard X-ray iron line, but
the paper does not quantify the energy and intensity of the feature.

To crudely test the plausibility of a wind fluid origin for the
Fe He$\alpha$ emission in these other starbursts we consider
NGC 253 ($L_{\rm Fe He\alpha} = (5.3^{+3.9}_{-2.8}) \times 10^{37} \ergps$, 
\citealt{weaver02}), 
Arp 299 ($L_{\rm Fe He\alpha} = (9.5\pm{4.5}) \times 10^{39} \ergps$,
\citealt{ballo04})
Arp 220 ($L_{\rm Fe He\alpha} = (1.2\pm{0.6}) \times 10^{40} \ergps$,
\citealt{iwasawa05}) and Mrk 273 
($L_{\rm Fe He\alpha} = (9.9^{+8.6}_{-8.1}) \times 10^{40} \ergps$, 
\citealt{balestra05}). We will use the $\lambda = 8$ -- 1000 $\mu$m IR 
luminosities of these galaxies, taken from
the data in \citet{sanders03}, 
as proxies for the star formation rate $\cal{S}$, making no
correction for the fraction of the IR luminosity contributed by any AGN. 
The $L_{\rm Fe He\alpha}$
luminosities given above are based on the 6.7 keV line photon fluxes
quoted in the appropriate paper and scaled to the adopted distances
given below.

\citet{weaver02} found that the hard X-ray emission in NGC 253 
(D=2.6 Mpc, $L_{\rm IR} = 1.9 \times 10^{10} \Lsol$) 
comes from a 300 pc by 60 pc region associated with the nuclear
starburst region and gas torus. If we interpret this as a disk of
that diameter and height then the equivalent 1-dimensional effective
radius is $R_{\star} = 126$ pc. Scaling M82's observed 
Fe He$\alpha$ emission by $({\cal{S}}_{\rm NGC 253}/{\cal{S}}_{\rm M82})^{2} 
\times (R_{\rm \star, M82}/R_{\rm \star, NGC 253})$ we obtain a predicted
Fe He$\alpha$ luminosity associated with the starburst 
of $L_{\rm Fe He\alpha} \sim 4.4 \times 
10^{37} \ergps$, which is consistent with the observed value given above.

Arp 299 (D=48.5 Mpc) is a merging system with a total IR luminosity of
$L_{\rm IR} = 7.7 \times 10^{11} \Lsol$, although the Fe He$\alpha$
emission is only associated with the eastern nucleus (IC 694).
Arp 220 (D=74 Mpc) and Mrk 273 (D=158 Mpc) have comparable IR luminosities of 
$L_{\rm IR} = 1.3 \times 10^{12} \Lsol$. As the sizes and geometries of the 
starburst regions in these galaxies are not well established we will
ignore the $R_{\star}$ scaling and scale the observed M82 Fe He$\alpha$
luminosity only by the square of the ratio of the IR luminosities.
For Arp 299 the predicted Fe He$\alpha$ luminosity is 
$L_{\rm Fe He\alpha} \sim 3.0 \times 10^{40} \ergps$, a few times higher
than the observed value of $\sim 10^{40} \ergps$.
For Arp 220 and Mrk 273 the predicted luminosity is 
$L_{\rm Fe He\alpha} \sim 8.5 \times 
10^{40} \ergps$. This value is
 comparable to the (poorly-constrained) Fe 
He$\alpha$ luminosity of Mrk 273, and several times larger than the observed
Fe He$\alpha$ luminosity of Arp 220.
As we would expect the effective size of the starburst
regions in these objects to be larger than that of M82 we would expect the
luminosity to somewhat less than the predictions given above, so these
predictions are in reasonable agreement with the observations.

Such back of the envelope calculations certainly
do not prove that the Fe He$\alpha$
emission seen in these objects arises from the same processes as in
M82. They do demonstrate that there is some hope of being able to test
and explore this model in objects other than M82 with current X-ray data
from the {\it Chandra} and {\it XMM-Newton} X-ray Observatories.

In the longer term the superior spectral resolution and sensitivity
of the calorimeter instrument on the International X-ray Observatory 
will allow routine detection of hard X-ray line emission from
a much larger sample of star forming galaxies. Furthermore, with
an energy resolution of $\Delta E \sim 4$ eV at $E=6.7$ keV it will
allow us to test the assumption that these lines arise in collisionally
heated gas.

\section{Conclusions}
\label{sec:conclusions}

We have measured the net energy efficiency of supernova and stellar wind
feedback (the thermalization efficiency $\epsilon$)
in the starburst galaxy M82 and the degree of central
mass-loading $\beta$ of its superwind by comparing a large suite of
1 and 2-dimensional hydrodynamical models to a set of observational
constraints derived from hard X-ray observations of the starburst
region (the fluxes of the He$\alpha$ and Ly$\alpha$-like lines 
of S, Ar, Ca and Fe, along
with the total diffuse $E=2$ -- 8 keV X-ray luminosity). 

The observational constraints are based on a combination of the
best currently available observations of the central 500 pc of
M82 with both the {\it Chandra} and {\it XMM-Newton} X-ray 
Observatories. We combine the results presented in \citet{strickland07} 
on the diffuse hard X-ray Fe line emission and the diffuse $E=2$ -- 8 keV 
hard X-ray luminosity with new results 
(this paper) on the S, Ar and Ca lines. By simultaneously applying
a large number of observational constraints (we used eight different
line fluxes or upper limits, 
an upper limit on the broad-band X-ray luminosity and a physical requirement
that the flow not be strongly radiative) to the simulated starbursts
we have been able to constrain
the physical properties of the starburst region more tightly than by
using more standard X-ray diagnostics.

We modeled the physical properties of the SN-heated and enriched
plasma within an assumed disk-like starburst region and the flow 
of gas outside this region out to a radius of 500 pc using either
an appropriately scaled version of the 1-dimensional \citet{chevclegg} 
analytical superwind solution, or with a full hydrodynamical 
simulation using the VH-1 hydrodynamics code in 2-dimensional cylindrical 
symmetry. We demonstrated a method for correcting the 1-dimensional radial
CC model so as to apply to a disk-like starburst geometry with reasonable
accuracy. Although this introduces errors of $\la 30$\% in the individual 
predicted hard X-ray line fluxes compared to the full 2-D hydrodynamical
solution, such errors have little effect on the derived constraints
on $\epsilon$ and $\beta$ as the uncertainties in the observed line
fluxes can be larger than this.  We also considered the
issue of the applicability of the adiabatic CC wind solution and
find that for the study of the volume-filling gaseous component of the
M82 superwind within individual star clusters and over the starburst 
region as a whole the use of adiabatic models is valid.

We considered a broad range of plausible parameters for the M82 starburst,
varying the age and mode of star formation, the starburst region size
and geometry, and supernova metal yields. We further considered a 
correction accounting for the fraction of
massive stars that occur in high density regions where high radiative energy
losses will preclude their making any contribution to the superwind, 
which we term the participation fraction $\zeta$.

We find that over all these varied input parameters (we
calculated 663 full hydrodynamical simulations using different model 
parameters, and 4615 models using the
scaled CC model), those models that
do satisfy the existing observational constraints all have medium to high
SN+SW thermalization efficiencies ($30$\% $\le \epsilon \le 100$\%)
and have mildly centrally
mass-loaded wind fluids (generally $1.5 \le \beta \le 2.5$,
although in a few cases $\beta = 1$ is allowed).

Our results are ultimately based on the measured Fe He$\alpha$~emission
of the SN heated and enriched plasma driving M82's wind. As such they
represent the first direct measurement of the efficiency of SN feedback and
mass-loading of the wind fluid in a starburst galaxy. Future observations
with higher spectral resolution (\eg with {\it ASTRO-H} and {\it IXO})
will be required to test whether the observed line emission truly
arises from a collisional plasma in or close to ionization equilibrium.

That the majority of the models that successfully match our imposed 
hard X-ray observational constraints also predict thermal pressures
in the starburst region 
($7.0 \la \log P_{\rm c}/k \la 7.5$)
that are consistent with prior independent estimates of the starburst
region pressure increases our
confidence in the validity of our method and results.

The allowed values of the SN and SW thermalization efficiency and
the mass loading factor correspond to a temperature for the 
metal-enriched plasma
within the starburst region of 30 -- 80 million Kelvin, 
a mass flow rate of the wind
fluid out of the starburst region of
$\Mdot_{\rm tot} \sim 1.4$ -- $3.6 \Msol \pyr$
and imply a terminal
velocity of the wind (in free expansion) in the range $v_{\infty} = 1410$
-- $2240 \kmps$. This velocity is considerably larger than the
gravitational escape velocity from M82 ($v_{\rm esc} \la 460 \kmps$)
and the velocity of the \halpha~emitting clumps and filaments
in the inner 2 kpc of M82's wind ($v_{\rm H\alpha} \sim 600 \kmps$). 
Reducing the velocity of the wind fluid once \emph{outside} the starburst
region to the escape velocity would require either unrealistically
large mass loading ($\Mdot_{\rm extra} \sim 13$ -- $240 \Msol \pyr$) 
or excessive energy losses ($\sim 90$ -- 98\%) that must also avoid
raising the luminosity of the wind.
Thus the hot gas driving M82's wind will probably escape the galaxy's
potential well. This result also adds support to the idea that the
long term fate of gas within superwinds is phase-dependent.

We show that the luminosity of the hard X-ray line emission produced
by the wind fluid in a starburst region scales with the star formation
rate ${\cal{S}}$ and starburst size $R_{\star}$ 
as $L_{\rm HX} \propto {\cal{S}}^{2}/R_{\star}$. This should serve
to distinguish this emission from the other sources of both
soft and hard X-ray emission in star forming galaxies that scale
linearly with the star formation rate. We show that existing measurements
of Fe He$\alpha$ emission from other actively actively star-forming galaxies
(NGC 253, Arp 299, Arp 220, Mrk 273),
previously interpreted as ionized reflection from their AGN, are
also consistent with our prediction for the wind fluid given the
existing observational uncertainties.

Drawing on our results for M82 we have also presented a simple
prescription  for implementing
starburst-driven superwinds in cosmological models of galaxy formation
and evolution that more accurately represents the energetics of the
hot metal-enriched phases than do existing recipes in the literature.

\acknowledgements We thank the anonymous referee for their careful
reading of this manuscript and their helpful recommendations.
This work was funded through grant 
NNG05GF62G of the NASA Astrophysics Theory Program. It is 
based on observations obtained with 
the {\it Chandra} X-ray Observatory, which is 
operated by the Smithsonian Astrophysical Observatory on behalf of NASA, and 
{\it XMM-Newton}, an ESA science mission with instruments 
and contributions directly funded by ESA Member States and NASA.

\appendix

\section{The conditions under which the adiabatic solution becomes invalid}
\label{app:radiative_case}

The basic adiabatic CC model is a function of only three parameters: 
$\Edot_{\rm tot}$, $\Mdot_{\rm tot}$, and $R_{\star}$. \citet{silich04}
were the first to point out that for certain values of these parameters,
in particular small $R_{\star}$ and/or high $\Mdot_{\rm tot}$, the
total luminosity predicted by the CC model was comparable to the net
energy injection rate and that the analytical CC solution was no longer 
self-consistent as it assumed energy losses to be negligible.
They developed a self-consistent radiative solution,
later tested with numerical hydrodynamical simulations \citep{tenorio-tagle07}.

There in not an exact and easily evaluated function 
that may be used to assess whether the adiabatic assumption 
is valid for any particular
combination of $\Edot_{\rm tot}$, $\Mdot_{\rm tot}$, and $R_{\star}$.
Nevertheless, \citet{silich04} state that
the threshold between the adiabatic and the
fully radiative solution occurs when the radiative energy losses 
per unit volume exceed $\sim 30$\% of the energy injection rate 
per unit volume.
 \citet{tenorio-tagle07} show that this is most likely
to occur at the very center of the star cluster or starburst region.

In terms of the variables defined in \S~\ref{sec:theory}, the adiabatic
solution becomes invalid when
\begin{equation}
\frac{ \chi_{\rm H} \, (1+\chi_{\rm H}) \, 
  \rho_{\rm c}^{2} 
  \, \Lambda(T_{\rm c, Z_{\rm WF}}) }{2 m_{\rm H}^{2}} \ga
  \frac{ 0.3 \, \Edot_{\rm tot}}{V_{\star}}.
\label{equ:app:radsol}
\end{equation}

Equivalently, if 
\begin{equation}
\frac{ 0.74 \, \zeta \, \beta^{3} \, \Mdot_{\rm SN+SW}^{3} \,
  \Lambda(T_{\rm c, Z_{\rm WF}}) 
  }{ m_{\rm H}^{2} \, \epsilon^{2} \, \Edot_{\rm SN+SW}^{2} \, R_{\star}} \ga
  1
\label{equ:app:radthresh}
\end{equation}
then radiative losses are high and the CC model is invalid. Note the 
strong dependence on the mass injection rate and any mass loading, 
and inverse dependence on the energy injection rate, thermalization
efficiency and effective radius. The higher the mean 
energy per particle, or the larger the starbursting  region is,
 the less-likely significant radiative losses
become.
 
\subsection{Individual star clusters}

\citet{silich04} and 
\citet{tenorio-tagle07} provide examples of star clusters with properties
such that radiative losses are severe. However, the mass and energy
injection rates they generally 
use have significantly lower values of the energy per particle
(and hence higher radiative losses for a given assumed metal abundance) 
than what we consider appropriate for clusters in a M82-like starburst.

Equation~\ref{equ:app:radthresh} may be rewritten as an explicit function
of the cluster mass. We rewrite the energy and
mass injection rates as a function of the star cluster mass assuming
in this case an instantaneous starburst model, such that 
$\Edot_{\rm SN+SW} = j M_{\rm 6}$ and 
$\Mdot_{\rm SN+SW} = k M_{\rm 6}$, where $M_{\rm 6}$ is the total
cluster mass in units of $10^{6} \Msol$ (again using 
a Salpeter IMF between the 
lower and upper mass limits of 1 -- $100 \Msol$). Note that the value
of $j$ and $k$ 
are a functions of the age of the cluster, and are largest in magnitude
for a cluster age of $t \sim 4$ Myr in the Starburst99 models. 
The critical cluster mass above which cooling exceeds 30\% of the 
energy injection and the CC model is no longer valid is then
\begin{equation}
M_{\rm 6} \ga \frac{ m_{\rm H}^{2} \, \epsilon_{\rm cl}^{2} \, 
  j^{2} \, R_{\star, \rm cl}}{ 0.74 \, \zeta_{\rm cl} \, 
  \beta_{\rm cl}^{3} k^{3} \, \Lambda(T,Z_{\rm WF})}.
\label{equ:app:m6_v1}
\end{equation}

For the purposes of illustration we consider a cluster of age $t = 5$ Myr,
for which $j M_{\rm 6} = 3.7 \times 10^{40} \ergps$ 
and $k M_{\rm 6} = 2.3 \times 10^{-2} \Msol \pyr$. For unit $\epsilon$,
$\beta$, and $\zeta$ the central temperature in the CC model would be
$T_{c} \sim 7.5 \times 10^{7} \K$ (equivalent to $v_{\infty} = 2250 \kmps$). 
Assuming $Z_{\rm SN+SW} = 5 \Zsol$ the
cooling function is 
$\Lambda(T,Z_{\rm WF}) = 5.54 \times 10^{-23} \erg$ cm$^{3}$ $\ps$.

The massive star clusters observed by \citet{mccrady07} in the
nuclear starburst region of M82 have half-light
radii of between 1 -- 3 pc. We adopt a typical cluster radius of 
$R_{\star, \rm cl}$ = 2 pc. Substituting these values into 
Equation~\ref{equ:app:m6_v1} we find
\begin{equation}
M_{\rm 6} \ga 1.9 \times 10^{2} \, 
  \frac{\epsilon_{\rm cl}^{2}}{\zeta_{\rm cl} \, \beta_{\rm cl}^{3}} \,
  \left( \frac{5.54 \times 10^{-23}}{\Lambda(T,Z_{\rm WF})} \right) \, 
  \left( \frac{R_{\star, \rm cl}}{2 \pc}\right).
\label{equ:app:m6_v2}
\end{equation}
For unit thermalization efficiency and no additional
mass loading then
no realistic cluster will lie in the radiative regime. Significant
mass loading within the cluster (\eg $\beta_{\rm cl} =5$) or 
low thermalization efficiency ($\epsilon_{\rm cl} = 0.1$) are required to
reduce $M_{\rm 6}$ to of order unity for a $t = 5$ Myr old cluster.

An older cluster is even less likely to be radiative. From the
Starburst99 models we find that at an age of
20 Myr the mass injection from stellar winds has declined dramatically
yielding a net mass injection rate a factor 5 lower than at an age of 5 Myr. 
The energy injection rate is also lower, but only by a 2.5. At this age
even clusters of mass a few times $10^{7} \Msol$ are still adiabatic even
if $\beta_{\rm cl} =5$ or $\epsilon_{\rm cl} = 0.1$.

\subsection{Starburst regions as a whole}

We rewrite the energy and mass injection rates as a function of the
star formation rate $\cal{S}$, so that $\Edot_{\rm SN+SW} = a \cal{S}$ and 
$\Mdot_{\rm SN+SW} = b \cal{S}$. If we use the values given in
Table~\ref{tab:burst} for constant star formation with an age of 30 Myr, 
again assuming a metal abundance of $Z_{\rm SN+SW} = 5 \Zsol$, we
find that the star formation rate must exceed
\begin{equation}
{\cal{S}} \ga 1.4 \times 10^{3} \, 
  \frac{\epsilon^{2}}{\zeta \, \beta^{3}} \,
  \left( \frac{5.85 \times 10^{-23}}{\Lambda(T,Z_{\rm WF})} \right) \, 
  \left( \frac{R_{\star}}{100 \pc}\right).
\label{equ:app:sfr_crit}
\end{equation}
before the adiabatic CC solution becomes invalid. 

For M82 the
star formation rate is ${\cal{S}} \sim 4$ -- $6 \Msol \pyr$ for the Salpeter
IMF assumed, and the starburst region has effective radius 
of $R_{\star} \sim 100$ pc
at smallest (and more likely $R_{\star} \sim 300$ pc). Thus the
CC model is valid unless mass loading is significant 
$\beta \ga 5$) or thermalization is inefficient $\epsilon  \la 0.1$, or
some combination of the two. Note that a low value for the
participation fraction $\zeta$ will 
reduce the tendency of the starburst region to be radiative.

We conclude that under most, but not all, conditions the
adiabatic \citet{chevclegg} model is appropriate for M82's nuclear starburst.

\section{Corrections to the CC model due to a non-spherical starburst region}
\label{app:non_spherical}

It is possible to use the spherical analytical CC model to predict the 
wind-fluid properties and emission from a disk-like starburst of 
diameter $d_{\star}$
and total height $h_{\star}$ by applying modest numerical corrections
that are a fixed function of the geometry alone. As long as the
adiabatic assumption is valid these correction factors
do not depend on the starburst energy or mass injection rates, and
are thus independent of the thermalization efficiency $\epsilon$
and the mass loading factor $\beta$.

In the analytical CC model the 
surface area of the starburst region $S_\star$ controls the gas 
density within
the starburst region by limiting the rate at which matter can flow out of
it, as described in \S~\ref{sec:ccmodel} (see in particular
Equ.~30 of \citealt{canto00}). 
Note that other parameters, in particular the central temperature and 
the terminal velocity of 
the wind are independent of the starburst geometry.
Thus, for a given fixed energy and mass injection rate,
the central density and pressure in a disk-like starburst of surface
area $S_{\star} = \pi d_{\star} \, (d_{\star}/2 + h_{\star})$ 
should be very similar or identical to values in a spherical starburst
of the same surface area $S_{\star}$. 

We tested whether this is indeed the case using the VH-1 hydrodynamical 
code run in 2-D cylindrical symmetry as described in \S~\ref{sec:hydromodel}. 
This code reproduces the standard spherical CC solution to within an
accuracy of 0.1\% for an assumed spherical starburst. For a disk-like
starburst of diameter  $d_{\star} = 750$ pc and 
$h_{\star} = 105$ pc (\S~\ref{sec:theory:applying:sbsize}) a spherical
starburst of equivalent surface area has $R_{\star} = 300$ pc. 

We find that the central density and pressure predicted by VH-1
for this disk-like starburst geometry differs from the values it
predicts for a $R_{\star} = 300$ pc spherical
starburst by only 3.1\%. Adopting another disk-like geometry with the
same surface area ($d_{\star} = 682$ pc and 
$h_{\star} = 200$ pc) yields central density and pressures within
4.2\% of the spherical solution. For the other disk-like starburst
geometry suggested by observations of M82, with $d_{\star} = 1015$ pc and 
$h_{\star} = 140$ pc the 2-D simulations predict central densities
and pressures within 3.7\% of the spherical starburst with equivalent
surface area ($R_{\star} = 405$ pc). These simulations demonstrate that
we can accurately predict the gas properties \emph{within} 
a disk-like starburst
region using the spherical CC model of equivalent starburst surface area.

\begin{figure*}[!t]
\epsscale{0.9}
\plotone{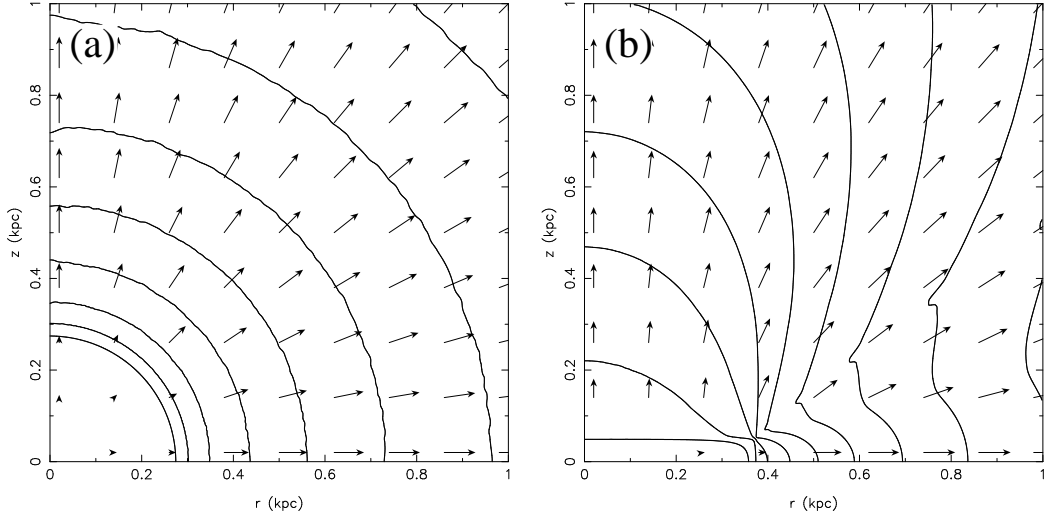}
  \caption{Gas number density and 
  velocity vectors in two hydrodynamical simulations
  of a starburst in (a) a spherical starburst of radius $R_{\star} = 300$ pc
  and (b) a disk-like starburst with radius $r_{\star} = 375$ pc and total
  thickness $h_{\star} = 105$ pc. The central gas density and pressure
  is the same for these configurations. The highest density contour shown
  corresponds to $\log n = -1.00$ and further contours correspond to
  successive decreases in $\log n$ in steps of 0.25. 
  }
  \label{fig:2dsim}
\end{figure*}

The flow pattern exterior to the starburst region is also influenced by the
geometry of the starburst region. 
 A disk with $d_{\star} > h_{\star}$ will weakly collimate
the flow along the minor axis and give the superwind a preferred direction
even in the absence of external collimation by the ISM.
In Fig.~\ref{fig:2dsim} we illustrate this
by plotting the density and velocity fields for equivalent spherical
and disk-like starbursts.

For our purposes the main effect is that
the resulting non-spherical flow alters the total volume emission integral
within any fixed region that is larger than the starburst region.
Within the starburst region itself this
effect is not particularly significant when evaluating total gas
masses or emission integral given the relative uniformity of the 
density and pressure. However, outside and above a disk-like starburst region
the gas density drops off with increasing height more slowly that
the $\rho \propto R^{-2}$ predicted by the CC model. 

Calculating the correct emission integral for a disk-like starburst
based the results of the spherical CC model thus entails some
geometry-based corrections because of 
the different starburst region volumes and secondly due to the non-spherical
flow pattern outside the starburst region. The first correction is
trivial to evaluate analytically. The second correction must be determined
from a full hydrodynamical simulation of the specific disk-like starburst,
and further depends on the size of region $R_{\rm obs}$ 
we wish to evaluate the emission integral over.

Our procedure for calculating the gas properties and total 
emission of disk-like starburst regions using the spherical CC
model is as follows:
\begin{enumerate}
\item For a given starburst geometry evaluate the surface area $S_{\star}$,
  and run a simulation using the CC model using 
  the effective starburst radius $R_{\star} = (S_{\star}/4\pi)^{1/2}$.
  This accurately predicts the values of fluid variables such as density,
  temperature and pressure \emph{within} the starburst region.
\item Macroscopic variables, in particular the volume emission integral,
  require further geometrical corrections to make the emission integral
  predicted using the spherical CC model, $EI_{\rm CC}$, match the
  emission integral found in a full multi-dimensional simulation of a
  non-spherical starburst region $EI_{\rm nonsph}$.
\item The volume of the starburst region in the spherical model is larger
  than the volume of the disk-like starburst region, so the volume
  emission integral predicted by the CC model within some radius
  $R_{\rm obs} \ge R_{\star}$ must be scaled by a factor $K_{V}$ equal to the
  ratio of the starburst region volumes:
  \begin{equation}
  K_{V} = \frac{\pi r_{\star}^{2} h_{\star}}{(4\pi/3) R^{3}_{\star}}.
  \label{equ:kvol}
  \end{equation}
\item A full hydrodynamical simulation for a given disk-like starburst
  must be run to evaluate the 
  correction $K_{\rm nonsph} = EI_{\rm nonsph} / (K_{V} \, EI_{\rm CC})$ 
  to the emission integral within the 
  chosen $R_{\rm obs}$ 
  due to the non-spherical flow pattern manifest outside the starburst region.
  We use $R_{\rm obs} = 500$ pc to match the region from which
  the M82 nuclear region spectra analyzed in
  Paper I were extracted. Simulations of different disk-like starburst regions
where we altered the thermalization efficiency and mass loading 
demonstrated that the correction $K_{\rm nonsph}$ 
to the total emission integral
is a fixed function of the geometry, and is independent of the
energy and mass injection rates.
\item The temperature-dependent differential emission measure for the 
  scaled CC model $EI_{\rm SCC}(T)$
  is calculated using the factors discussed above: 
  $EI_{\rm SCC}(T) = K_{V} \times K_{\rm nonsph} \times EI^{\rm CC}(T)$.
  Combining this with the temperature-dependent cooling functions and summing
  over all temperatures gives the final predicted luminosity for the chosen
  energy band or line.
\end{enumerate}

For $R_{\rm obs} = 500$ pc and a disk-like starburst with  
$d_{\star} = 750$ pc and $h_{\star} = 105$ pc, the primary
geometry we apply to M82, the non-spherical emission
integral correction factor $K_{\rm nonsph} = 1.64$. For the
alternate disk-like starburst
geometry with $d_{\star} = 1015$ pc and $h_{\star} = 140$ pc
the correction factor is $K_{\rm nonsph} = 1.52$. 
For the compact starbursting disk
with $d_{\star} = 300$ pc and $h_{\star} = 60$ pc
the correction factor is $K_{\rm nonsph} = 1.59$.

With these scalings the spherical CC model can be used to predict to high
accuracy the total emission integral from the wind
fluid within $R_{\rm obs}$ for a disk-like 
starburst. This method is less accurate at reproducing the luminosity
predicted by a full 2-dimensional hydrodynamical simulation because
non-spherical flow pattern leads to subtle differences in the shape of
the differential emission integral EI(T) compared to a spherical CC solution.

We compared the luminosities predicted by a small suite of hydrodynamical 
models with different disk-like starburst geometries to the results from 
the scaled CC
analytical model in order to assess the significance of this effect.
The error introduced by the effect is small in terms of computing the total 
cooling rate ($< 2$\% for the Sutherland \& Dopita cooling curve), or even 
the the X-ray luminosity over the broad 0.3 -- 2.0 keV or 2 -- 8 keV energy
ranges ($\la 10$\%). For the high energy lines we are interested in because
of their temperature sensitivity the effect is larger, and the
associated errors in predicted line luminosity can be as large as 
$\sim 30$\%. 

Nevertheless these uncertainties do not compromise
the usefulness of the scaled CC model as they are comparable
to the observational uncertainties in the nuclear region
S, Ar and Ca luminosities of the well detected
lines from the {\it XMM-Newton} data, and less than the uncertainties 
in the diffuse Fe He$\alpha$ line luminosity derived from the
{\it Chandra} ACIS-S data. The relative effect of these uncertainties 
on the region of model parameter space matching our chosen constraints
can be seen in Fig.~\ref{fig:vh1_vs_scaledcc}.

\section{Estimating the participation fraction $\zeta$}
\label{app:zeta}

There are two main environments that in principle might reduce the 
fraction $\zeta$ of all  the SNe and stellar winds in a starburst region that
can contribute to powering a superwind.
The first, likely to to be the most significant but also hardest to estimate, 
is physical confinement and increased radiative
cooling of young supernova remnants 
and stellar winds occurring in regions of high ambient
gas density such as molecular clouds. 
The second is that radiative losses may be high for those
SNe and stellar winds associated with massive stars
in massive and compact star clusters (particularly super star clusters),
where the adiabatic CC solution is no longer valid.

There are some observational findings that are suggest that some fraction
of the SNe in M82's nuclear starburst occurred within regions of gas
number density $n \ga 10^{3} \pcc$ \citep{kronberg00,chev01,seaquist07},
although there are problems with this interpretation \citep[see][and 
references therein]{beswick06}. However, based on the total molecular
gas mass of $\sim 10^{8} \Msol$ within the starburst region, 
the fraction of the 
total starburst region volume occupied by such dense gas must be $\eta \la 0.1$
or lower \citep{lugten86,nakai87}. Massive stars will be born
in regions of high gas density, so that if confinement does occur the magnitude
of the drop in $\zeta$ may well be larger the the filling factor of the
dense gas. 
At present we can not rigorously quantify the effectiveness of
confinement, nor the fraction of stars that might be affected.
\citet{chev01} speculate that perhaps as much as 50\% of the SNe in M82
occur in dense regions, in which case the participation factor may be
as low as $\zeta \sim 0.5$.

It is possible to quantitatively estimate the role that over-cooling 
of the thermalized SN and stellar wind ejecta
in dense star clusters have on $\zeta$. Based on the discussion in 
Appendix~\ref{app:radiative_case} we conclude that only the compact 
clusters of radii a few parsecs and masses $M_{\rm rad} \ga 10^{6} \Msol$ 
(Salpeter IMF, mass limits 1 -- 100 $\Msol$) are likely to suffer significant 
radiative losses (and then only under certain conditions). 

For a star cluster mass function of the form 
$dN_{\rm cl} \propto M_{\rm cl}^{-\gamma} \, dM$ between the mass limits
$M_{\rm cl, min}$ and $M_{\rm cl, max}$ and where $\gamma = 2$ 
\citep{weidner06}, 
the fraction of the total
stellar mass in clusters with $M_{\rm cl} \ga M_{\rm rad}$ is
\begin{equation}
f = \frac{\ln M_{\rm cl, max} - \ln M_{\rm rad}}{\ln M_{\rm cl, max} - \ln M_{\rm min}}.
\end{equation}

We adopt a minimum cluster mass of 
$M_{\rm cl, min} = 5 \Msol$ and a maximum cluster mass of 
$M_{\rm cl, max} = 10^{7} \Msol$
\citep{weidner04b}. For $M_{\rm rad} = 10^{6} \Msol$ then $f = 0.16$. Thus
the magnitude of the reduction in $\zeta$ due to radiative cluster winds
is small, of order 15\%, even if SN and stellar wind
thermalization is inefficient, and/or there are
long-lived sources of additional mass loading in the massive star clusters.


\end{document}